\documentclass[10pt,a4paper,fleqn]{scrartcl}
\pdfoutput=1
\usepackage{CJK}
\usepackage[utf8]{inputenc}
\usepackage[english]{babel}
\usepackage[utf8]{inputenc}
\usepackage{placeins}
\usepackage[T1]{fontenc}
\usepackage{amsmath}
\usepackage{amsfonts}
\usepackage{amssymb}
\usepackage{graphicx}
\usepackage[left=2cm,right=2cm,top=2cm,bottom=2cm]{geometry}
\usepackage[version=3]{mhchem}
\usepackage{appendix}
\usepackage{multirow}
\usepackage{array}
\usepackage{siunitx}
\usepackage{subcaption}
\usepackage{cite}
\usepackage[font=footnotesize,margin=30pt]{caption}
\usepackage{chngcntr}
\usepackage{wrapfig}
\usepackage{url}
\usepackage[version=3]{mhchem}
\usepackage{braket}
\usepackage{authblk}
\usepackage{mathtools}
\usepackage{multicol}
\usepackage{hyperref}
\usepackage{cleveref}

\DeclareMathOperator{\Tr}{Tr}
\DeclareMathOperator{\sinc}{sinc}

\newcommand{\eqcolon}{\mathrel{\resizebox{\widthof{$\mathord{=}$}}{\height}{ $\!\!=\!\!\resizebox{1.2\width}{0.8\height}{\raisebox{0.23ex}{$\mathop{:}$}}\!\!$ }}}

\begin{document}

\title{Numerical Investigation of Photon-Pair Generation in Periodically Poled \textit{M}TiO\textit{X}O$_{4}$\\(\textit{M} = K, Rb, Cs; \textit{X} = P, As)}
\date{}

\author[1,2]{\large Fabian Laudenbach$^{*}$}
\author[3]{\large Rui-Bo Jin}
\author[2]{\large Chiara Greganti}
\author[1]{\large Michael Hentschel}
\author[2]{\large Philip Walther}
\author[1]{\large Hannes H\"{u}bel}
\affil[1]{\normalsize Security \& Communication Technologies, Center for Digital Safety \& Security,\protect \\AIT Austrian Institute of Technology GmbH, Donau-City-Str. 1, 1220 Vienna, Austria}
\affil[2]{\normalsize Quantum Optics, Quantum Nanophysics \& Quantum Information, Faculty of Physics, University of Vienna, Boltzmanngasse 5, 1090 Vienna, Austria}
\affil[3]{\normalsize Laboratory of Optical Information Technology, Wuhan Institute of Technology, Wuhan 430205, China \protect \\ \quad}
\affil[*]{\normalsize mail: fabian.laudenbach@ait.ac.at}

\maketitle

\begin{abstract}
We present a detailed numerical investigation of five nonlinear materials and their properties regarding photon-pair creation using parametric downconversion. Periodic poling of ferroelectric nonlinear materials is a convenient way to generate collinearly propagating photon pairs. Most applications and experiments use the well-known potassium titanyl phosphate (\ce{KTiOPO4}, \text{ppKTP}) and lithium niobate (\ce{LiNbO3}, \text{ppLN}) crystals for this purpose. In this article we provide a profound discussion on the family of KTP-isomorphic nonlinear materials, including KTP itself but also the much less common CTA (\ce{CsTiOAsO4}), KTA (\ce{KTiOAsO4}), RTA (\ce{RbTiOAsO4}) and RTP (\ce{RbTiOPO4}). We discuss in which way these crystals can be used for creation of spectrally pure downconversion states and generation of crystal-intrinsic polarisation- and frequency entanglement. The investigation of the new materials disclosed a whole new range of promising experimental setups, in some cases even outperforming the established materials \text{ppLN} and \text{ppKTP}.
\end{abstract}

\begin{multicols}{2}

\section{Introduction}

Photon-pair generation by spontaneous parametric downconversion (SPDC), a nonlinear process where a high-energy photon (mostly referred to as \emph{pump}) decays into two low-energy photons (\emph{signal} and \emph{idler}), is a convenient approach to realise heralded single-photon sources or quantum entanglement, vital for photonic quantum-information applications. In particular, SPDC in periodically poled nonlinear media  allows for collinear propagation and therefore high collection efficiencies. This technique is often referred to as quasi-phase-matching (QPM). Popular nonlinear crystals suited for periodic poling are \emph{potassium titanyl phosphate} (\ce{KTiOPO4}, \text{ppKTP}), \emph{lithium niobate} (\ce{LiNbO3}, \text{ppLN}) and \emph{lithium tantalate} (\ce{LiTaO3}, \text{ppLT}).

Most applications of photon-pair generation require a high spectral quantum purity which is equivalent to \emph{frequency-uncorrelated} photon pairs. However, in general, SPDC states do contain frequency-entanglement which, when one of the substates is traced out (i.e.\ by detection of one photon), puts the other one into a mixed state. In order to remove these correlations in the joint spectral distribution, many experiments use narrow bandpass filters~\cite{meyer2017filtering} or single-mode fibres~\cite{Guerriero2013} which only transmit the uncorrelated parts of the spectrum. This obviously comes with a drastic decrease in count rates and heralding efficiency since a large quantity of generated photons is discarded. As an alternative approach, the SPDC process can be carefully designed such that the generated SPDC state is \emph{a priori} frequency-uncorrelated, hence no filtering is required in order to obtain high purity~\cite{laudenbach2016modelling, mosley2008recipe, uren2006generation, jin2013widely, evans2010bright, eckstein2011highly, edamatsu2011photon, yabuno2012fourphoton, jin2013nonclassical, jin2014efficient, weston2016efficient, grice2001eliminating, mosley2008heralded}. This, however, is only feasible in a closed set of special cases, i.e.\ specific crystals allow for intrinsically pure SPDC states only at particular wavelength- and polarisation configurations. In addition, recently, three proposals have been suggested \cite{Dixon2013, Dosseva2016, graffitti2017pure} and implemented \cite{Chen2017} in which the joint spectral distribution is shaped by a particular custom fabrication of the poled crystals. Due to the early stage of development, it is difficult to evaluate the challenges in the fabrication process of identical crystals and therefore in the generation of independent identical photons.

As another highly interesting application of periodically poled crystals, we very recently demonstrated a way to generate photonic entanglement which is both very simplistic and efficient~\cite{laudenbach2016novel}. Up to now, efficient generation of collinear entangled photon pairs required either two periodically poled crystals~\cite{yoshizawa2003generation, clausen2014source, herbauts2013demonstration, pelton2004bright, huebel2007high} or, alternatively, one single crystal pumped from two directions simultaneously~\cite{shi2004generation, koenig2005efficient, kim2006phase, fedrizzi2007wavelength, hentschel2009three}. Our novel approach uses only one unidirectionally pumped \text{ppKTP} crystal which drastically simplifies the experimental setup without any compromise in count rates or entanglement visibility. While a comparable concept has been proposed and demonstrated for modal phase-matching on non-birefringent Bragg-reflection waveguides~\cite{horn2013inherent,kang2016monolithic}, our method benefits from the relatively easy and widespread fabrication of quasi-phase-matching structures (cost-efficient, insusceptible to fabrication errors) and the accurate spatial and spectral overlap of multiple decay channels (high brightness and entanglement visibility). It is based on the observation that a periodically poled crystal with a given grating constant, pumped with a given wavelength, can allow for two SPDC processes simultaneously, each exploiting its own quasi-phase-matching order\textemdash an effect we refer to as \emph{collinear double-downconversion} (CDDC). Unfortunately, this method only allows for a limited set of possible wavelength configurations, depending on the material properties of the respective nonlinear crystal. This means that, although there is a large set of possible realisations, the generated wavelengths cannot be chosen completely arbitrarily. However, the more nonlinear materials available, the greater is the repertory of possible configurations.

In the past we performed extensive numerical investigations of well-known optical materials to find out to what extent they support the above mentioned techniques of frequency-uncorrelated SPDC states~\cite{laudenbach2016modelling} and intrinsic entanglement \cite{laudenbach2016novel}. We found a large set of a priori pure downconversion processes in \text{ppKTP}, \text{ppLN} and \text{ppLT}, encompassing all polarisation configurations and various wavelength regimes. As for intrinsic entanglement by CDDC, we showed that \text{ppKTP} allows for numerous entangled-photon wavelengths in the near- and mid-infrared as well as in the telecom regime.

Lately, we started to extend the search for intrinsically pure SPDC states to other nonlinear materials, namely CTA (\ce{CsTiOAsO4}), KTA (\ce{KTiOAsO4}), RTA (\ce{RbTiOAsO4}) and RTP (\ce{RbTiOPO4})~\cite{jin2016spectrally}. Being isomorphic to KTP, these crystals all share similar properties~\cite{dmitriev1999handbook}. Like KTP they belong to the \emph{mm2} point group~\cite{sands1993dover}, they have a similar transparency range and nonlinear tensor. Moreover, being ferroelectric materials, they allow for periodic poling just as KTP does. In our recent article we presented the results of our numerical investigations of these materials, focusing on the important special case of frequency-degenerate and spectrally indistinguishable type-II downconversion with high spectral purity. For the present work, using our software \text{QPMoptics}~\cite{laudenbach2016qpmoptics}, we investigated the properties of the KTP isomorphs more thoroughly, searching for inherently pure states in all possible polarisation- and wavelength configurations, including the non-degenerate ones. Moreover, we studied their suitability for the above-mentioned approach for intrinsic-entanglement generation. In conjunction with \cite{laudenbach2016modelling} (\text{ppLN}, \text{ppLT}, \text{ppKTP}) the present article provides a comprehensive reference of a total of seven different optical media and in which way they can be used to generate pure photon states and/or intrinsic entanglement.

\section{Effective Nonlinear Coefficient}

\Cref{defftable} lists the approximate effective nonlinear coefficients for all kinds of polarisation configurations in five periodically poled nonlinear materials. The table illustrates the similarity of the five materials in terms of effective nonlinearity and hence downconversion efficiency.

\begin{table*} [h!]
\centering
\begin{tabular}{|>{\centering\arraybackslash}m{0.2in}|>{\centering\arraybackslash}m{0.75in}|>{\centering\arraybackslash}m{0.7in}|>{\centering\arraybackslash}m{0.7in}|>{\centering\arraybackslash}m{0.7in}|>{\centering\arraybackslash}m{0.7in}|>{\centering\arraybackslash}m{0.7in}|}
\hline
\multicolumn{2}{|c|}{\multirow{2}{*}{SPDC Type}} & \multicolumn{5}{c|}{Effective nonlinear Coefficient $|d_{\text{eff}}|~\left[\SI{}{\pico\metre\volt^{-1}}\right]$}  \\
\cline{3-7}
\multicolumn{2}{|c|}{} & ppKTP & ppCTA & ppKTA & ppRTA & ppRTP \\
\hline \hline
\multirow{2}{*}{0} & $o\longrightarrow o+o$ & 0 & 0 & 0 & 0 & 0  \\
\cline{2-7}
 & $e\longrightarrow e+e$ & $d^{33} \sim 9.5$ & $d^{33} \sim 11.2$ & $d^{33} \sim 9.6$ & $d^{33} \sim 9.8$ & $d^{33} \sim 9.6$ \\
\hline
\multirow{2}{*}{I} & $o\longrightarrow e+e$ & 0 & 0 & 0 & 0 & 0 \\
\cline{2-7}
 & $e\longrightarrow o+o$ & $d^{24} \sim  2.4$ & $d^{24} \sim  2.1$ & $d^{24} \sim  2.3$ & $d^{24} \sim  2.4$ & $d^{24} \sim  2.4$ \\
\hline
\multirow{2}{*}{II} & $o\longrightarrow o+e$ & $d^{32} \sim  2.4$ & $d^{32} \sim  2.1$ & $d^{32} \sim  2.3$ & $d^{32} \sim  2.4$ & $d^{32} \sim  2.4$ \\
\cline{2-7}
 & $e\longrightarrow o+e$ & 0 & 0 & 0 & 0 & 0 \\
\hline
\end{tabular}
\caption{Effective nonlinearity coefficients of five kinds of periodically poled crystals in all polarisation configurations, retrieved from the software \emph{SNLO v67}, developed by \emph{AS-Photonics, LLC}~\cite{asphotonics}. The supercripts of the nonlinear coefficients indicate the respective component of the nonlinear tensor $d$~\cite{powers2011fundamentals}. In all cases a collinear propagation along the crystal's $x$-axis is assumed. The letter $o$ denotes polarisation along the \emph{ordinary} (the $y$-) axis while $e$ denotes polarisation along the \emph{extraordinary} (the $z$-) axis. Note that all numerical values in the table are to be understood as approximation since they vary slightly with the involved wavelengths and the material's doping.}
\label{defftable}
\end{table*}

\section{Configurations of High Intrinsic Purity}

The Hong-Ou-Mandel visibility~\cite{hong1987measurement} of two photon states, represented as $\rho_{A}$ and $\rho_{B}$, is upper-bounded by the relation~\cite{mosley2008recipe}

\begin{align} \label{eq_VHOM}
V_{\text{HOM}} & \leq \Tr(\rho_{A} \rho_{B}) = \frac{P_{A}+P_{B}}{2} - \frac{\|\rho_{A}-\rho_{B} \|^{2}}{2} \notag \\
& \eqcolon \frac{P_{A}+P_{B}}{2} - \Delta ,
\end{align}
highlighting the role of quantum \emph{purity} $P \in [0,1]$ and \emph{distinguishability} $\Delta \in [0,1]$\footnote{The matrix norm in \eqref{eq_VHOM} indicates the Frobenius norm: $\| \rho \| = \sqrt{\Tr(\rho^{\dagger}\rho)}$.}. When the interfering photons originate from two identical sources, the purities coincide ($P_{A}=P_{B}\eqcolon P$) and the visibility simplifies to

\begin{align}
V_{\text{HOM}} \leq P - \Delta ,
\end{align}
where $\Delta \geq 0$ when the signal of one pair interferes with the idler of another and $\Delta=0$ when the corresponding photons of two pairs interfere (say, the two signal photons when the idler is used for heralding). Therefore, for identical heralded-photon sources the distinguishability does not play a role and the visibility depends on the purity only.

The purity of a quantum state $\rho$ is given by the trace of its square: $P=\Tr (\rho^{2})$. For a pure bipartite system $\ket{\Psi}_{AB}$, expressed in terms of a Schmidt decomposition

\begin{align}
\ket{\Psi}=\sum_{i} \sqrt{\lambda_{i} } \ket{i_{A}}\ket{i_{B}} ,
\end{align}
the purity of the substates is equivalent to the sum of the squared Schmidt coefficients:

\begin{align}
P_{A}=P_{B}=\sum_{i} \lambda_{i}^{2} \eqcolon \frac{1}{K} ,
\end{align}
where $K \geq 1$ is referred to the \emph{Schmidt number}, a convenient measure for the entanglement in a bipartite system. The above equation illustrates nicely the relation between purity and entanglement. In a highly entangled state there will be many non-zero coefficients $\lambda$ in the Schmidt decomposition, yielding a high degree of entanglement $K$ and therefore low purity $P$. Conversely, in a factorable state there is only one non-vanishing coefficient $\lambda=1$, yielding minimal entanglement $K=1$ and maximal purity $P=1$. Thus, a high spectral purity (as required for high interference visibilities) is equivalent to low spectral correlations of the interfering substates.

The Schmidt coefficients of a downconversion state $\ket{\Psi}$ can be obtained numerically by a singular-value decomposition (SVD)~\cite{mosley2008recipe} of the joint spectral amplitude (JSA) which is the product of the pump amplitude $\mu$ and the quasi-phase-matching (QPM) amplitude $\psi$:

\begin{align}
\text{JSA}=\mu(\omega_{s}+\omega_{i}) \psi(\omega_{s},\omega_{i}) .
\end{align}
The QPM amplitude reads

\begin{align} \label{QPMamplitude}
\psi=e^{i \Delta k_{m} L/2} \sinc \left( \frac{ \Delta k_{m} L }{2} \right) .
\end{align}
Here $L$ is the length of the nonlinear crystal and $\Delta k_{m}$ is the wave-vector difference

\begin{align}
\Delta k_{m} & = k_{p} - k_{s} - k_{i} - \frac{2 \pi m}{\Lambda} \notag \\
& =  \frac{\omega_{s}}{c} ( n_{p} - n_{s}) + \frac{\omega_{i}}{c}  ( n_{p} - n_{i}) - \frac{2 \pi m}{\Lambda} ,
\end{align}
where $n$ represents the refractive index, $\Lambda$ is the poling period of the crystal, $m$ is an odd integer referred to as QPM order, and we used energy conservation $\omega_{p}=\omega_{s}+\omega_{i}$. Quasi-phase-matching is achieved when the sinc function in \Cref{QPMamplitude} equals one, hence when $\Delta k_{m}=0$. Experimentally, we are not able to observe amplitudes directly. We therefore introduce the Schmidt number of the joint spectral \emph{intensity} $K_{\text{JSI}}$, obtained by an SVD of

\begin{align}
\text{JSI} = | \text{JSA} |^{2} .
\end{align}

The QPM condition $\Delta k_{m}=0$ needs to be met in order to generate a measurable output radiation with the desired wavelengths. On top of that, in order to allow for the generation of spectrally uncorrelated photon pairs, an SPDC setup needs to fulfil the \emph{group-velocity-matching} (GVM) condition. GVM can be achieved in two different ways: The first way is to match the group velocity of the pump photons with the group velocity of either the signal or the idler photons; in the second way the group velocity of the pump photons coincides with the average group velocity of the signal- and idler photons. The quality of GVM can be quantified by the dispersion parameter $D$~\cite{Kaneda2016}:

\begin{subequations} \label{dispersionparameter}
\begin{align}
D & =-\frac{\text{GD}_{p}-\text{GD}_{s}}{\text{GD}_{p}-\text{GD}_{i}} \qquad \text{or} \\
D & =-\frac{\text{GD}_{p}-\text{GD}_{i}}{\text{GD}_{p}-\text{GD}_{s}},
\end{align}
\end{subequations}
depending on whether the pump velocity is matched to the signal velocity (top line) or the idler velocity (bottom line). GD represents the respective group delay, hence the inverse group velocity in the optic medium. The two different ways of GVM yield

\begin{subequations}
\begin{align}
D & = 0 \qquad \text{if} \quad \text{GD}_{p}=\text{GD}_{s} \ \text{or} \ \text{GD}_{p}=\text{GD}_{i}, \\
D & = 1 \qquad \text{if} \quad \text{GD}_{p}=(\text{GD}_{s} + \text{GD}_{i})/2.
\end{align}
\end{subequations}
A dispersion parameter of $D=1$ corresponds to the QPM envelope amplitude $\psi$ being diagonally aligned in the $\lambda_s$-$\lambda_i$ plane. This allows for a circular-shaped JSA, as required by multiphoton experiments where signal and idler are supposed to be spectrally pure \emph{and} indistinguishable. Note, however, that maximal spectral indistinguishability of signal and idler to a certain degree compromises the state's purity and vice versa. This is due to the side lobes of the sinc function in the QPM amplitude $\psi$ which are maximally correlated due the diagonal orientation of $\psi$. This impairment limits the intrinsic purity achievable with $D=1$ to $P \sim 0.83$. On the other hand, when the GVM condition $D=0$ is met, $\psi$ is aligned horizontally or vertically in the wavelength plane, therefore carrying no spectral correlations. This gives rise to an intrinsic purity of $P > 0.96$. However, this configuration does not allow for spectral indistinguishability (even when the centre wavelengths are the same) since the pump amplitude $\mu$ and $\psi$ will never intersect orthogonally which will result in unequal signal and idler bandwidths. While $D=1$ is preferred for multiphoton experiments, GVM with $D=0$ is clearly favourable for heralded-photon sources where spectral indistinguishability between signal and idler is not a requirement.

\begin{figure*}
	\centering
\subcaptionbox{JSA$=\mu\psi$; $D\approx 1$, $P=V_{\text{HOM}} \sim 0.83$}
	[0.47\linewidth]{\includegraphics[width=0.47\linewidth]{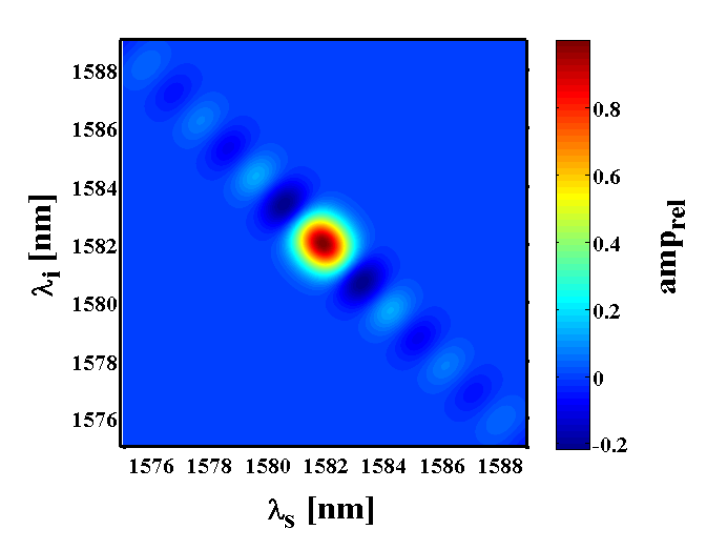}}
\subcaptionbox{$\mu+\psi$}
	[0.25\linewidth]{\includegraphics[width=0.25\linewidth]{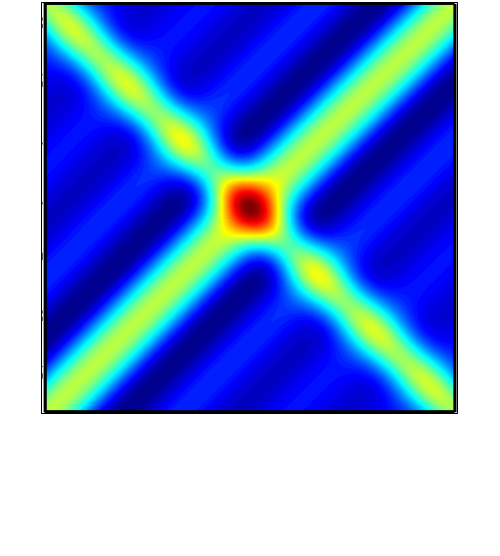}}
\subcaptionbox{$\psi$}
	[0.25\linewidth]{\includegraphics[width=0.25\linewidth]{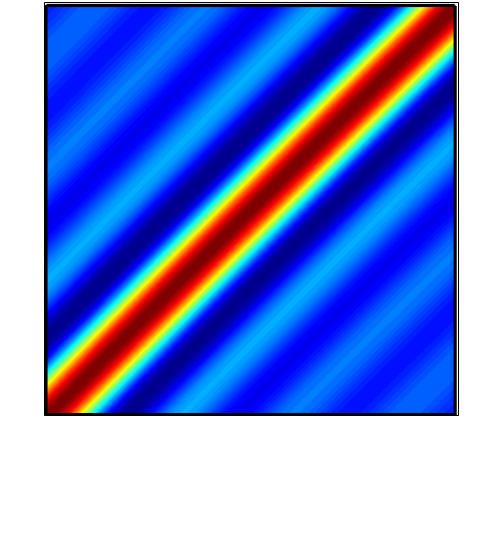}}
\subcaptionbox{JSA$=\mu\psi$; $D\approx 0$, $P \sim 0.96$, $V_{\text{HOM}} \sim 0.11$}
	[0.47\linewidth]{\includegraphics[width=0.47\linewidth]{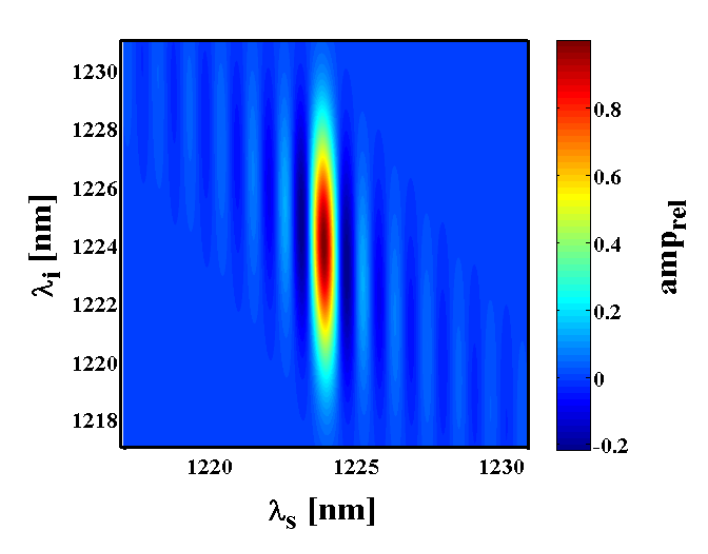}}
\subcaptionbox{$\mu+\psi$}
	[0.25\linewidth]{\includegraphics[width=0.25\linewidth]{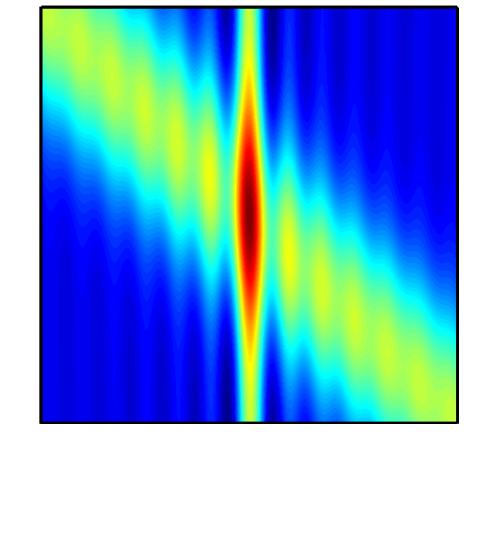}}
\subcaptionbox{$\psi$}
	[0.25\linewidth]{\includegraphics[width=0.25\linewidth]{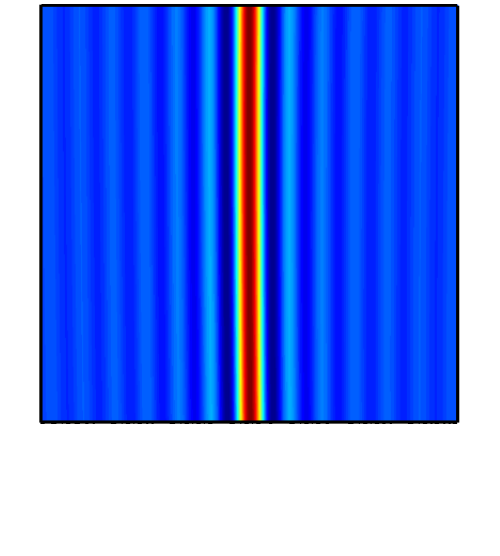}}
\caption{Group-velocity matching in two examples. Figure~(a) represents the downconversion $\SI{791}{\nano\metre}~(o) \longrightarrow \SI{1582}{\nano\metre}~(o) +\SI{1582}{\nano\metre}~(e)$ with the dispersion parameter $D\approx 1$, Figure~(d) corresponds to $\SI{612}{\nano\metre}~(o) \longrightarrow \SI{1224}{\nano\metre}~(o) +\SI{1224}{\nano\metre}~(e)$ with $D\approx 0$ (both in \text{ppKTP}). Signal and idler in Figure~(a) are spectrally fully indistinguishable which is allowed for by the orthogonal intersection of the pump- and QPM amplitude (b). This, however, requires the QPM amplitude $\psi$ to be aligned diagonally in the wavelength plane (c) with the side lobes of the sinc function being maximally spectrally correlated. This limits the maximal purity for spectrally indistinguishable SPDC states to $P=V_{\text{HOM}} \sim 0.83$. On the other hand, when the QPM amplitude $\mu$ is aligned more vertically (f) the side lobes are less spectrally correlated which allows for a higher purity of $P>0.96$, but for the price of undermined indistinguishability and therefore signal-idler-interference visibility $V_{\text{HOM}} \sim 0.1$. (The plots of $\mu+\psi$ (b,d) only serve the purpose to illustrate the composition of the JSA and do not correspond to an actual physical quantity; the illustrations of the QPM amplitude $\psi$ (c,f) correspond to the same colour mapping as the ones of the JSA (a,c).)}
\label{791and625}
\end{figure*}

The two ways of group-velocity matching are illustrated via examples in \Cref{791and625}. The subfigures~(a), (b) and (c) correspond to the type-II downconversion $\SI{791}{\nano\metre}~(o) \longrightarrow \SI{1582}{\nano\metre}~(o) +\SI{1582}{\nano\metre}~(e)$ in \text{ppKTP} with
 $D=-(\SI{6.027}{ps/mm} - \SI{5.880}{ps/mm} )/(\SI{6.027}{ps/mm} - \SI{6.175}{ps/mm})=0.993$. This configuration generates fully spectrally indistinguishable photons with a compromised purity of maximally $P \sim 0.83$. The subfigures~(d), (e) and (f) correspond to the process $\SI{612}{\nano\metre}~(o) \longrightarrow \SI{1224}{\nano\metre}~(o) +\SI{1224}{\nano\metre}~(e)$ in \text{ppKTP} with $D=-(\SI{6.209}{ps/mm} - \SI{6.208}{ps/mm} )/(\SI{6.209}{ps/mm} - \SI{5.903}{ps/mm})=-0.003$ and a purity of $P \sim 0.96$. Due to the differing bandwidth signal and idler are easily distinguishable which is a drawback for multiphoton experiments but irrelevant for heralding purposes.

\begin{figure*}
\centering
\includegraphics[width=.94\linewidth]{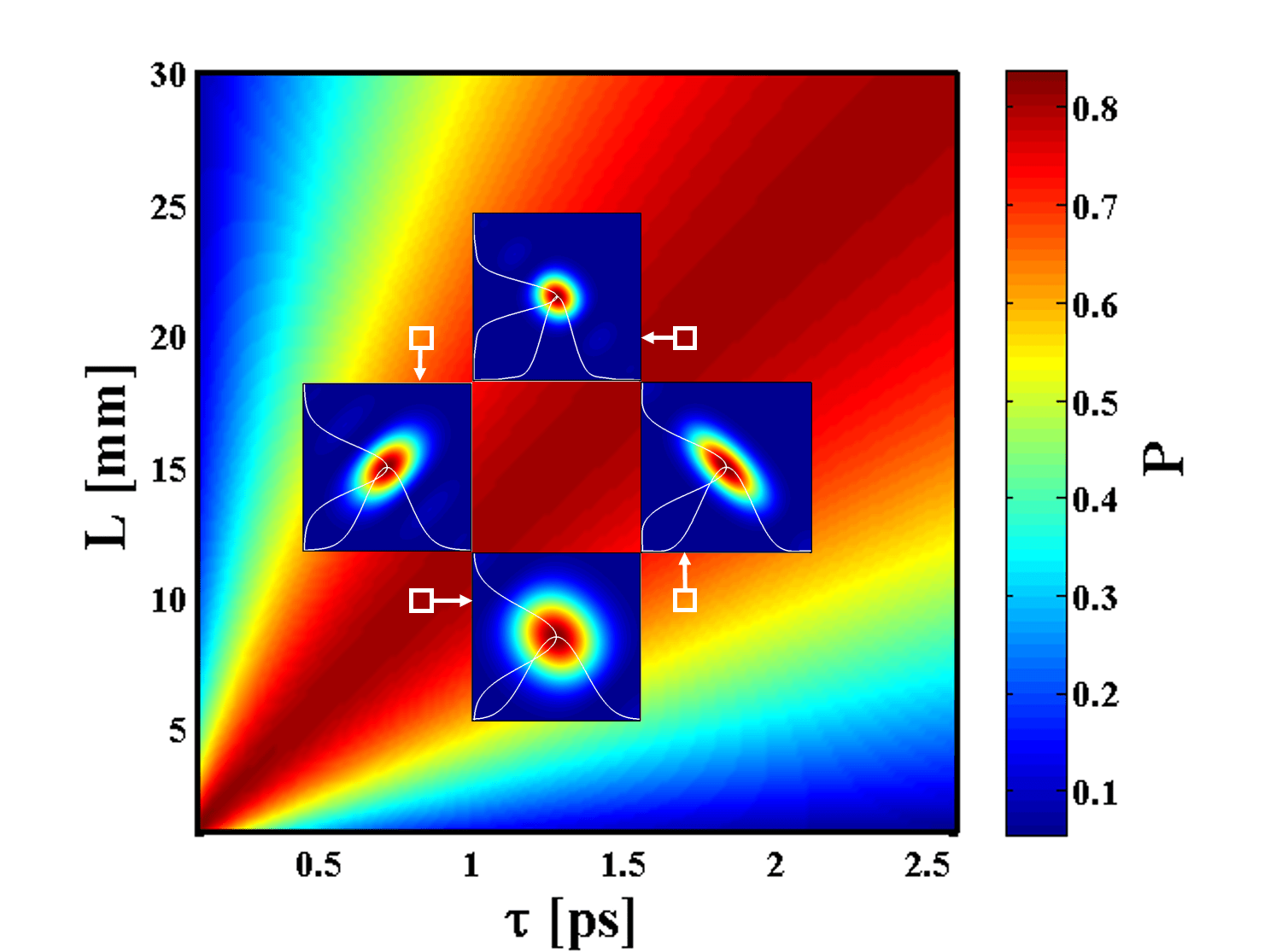}
\caption{Purity vs.\ Gaussian-pulse duration $\tau$ and crystal length $L$. The example corresponds to the downconversion $\SI{791}{\nano\metre}~(o) \longrightarrow \SI{1582}{\nano\metre}~(o) +\SI{1582}{\nano\metre}~(e)$ in \text{ppKTP}. The inset figures depict the joint spectral intensity distribution corresponding to particular pairs of $\tau$ and $L$. The white curves in the insets illustrate the spectral intensity distribution of signal (horizontal axis) and idler (vertical axis). Although signal and idler are spectrally indistinguishable for any pair of $\tau$ and $L$, a high intrinsic purity (i.e.\ frequency-uncorrelated daughter photons) can only be obtained by mutual matching of the two. A short-pulsed (and therefore broadband) laser pumping a long crystal will generate frequency-correlated spectra (left inset); a long-pulsed (hence narrowband) laser in a short crystal will produce anti-correlated signal and idler (right inset). The top and bottom inset represent spectrally uncorrelated SPDC states, achieved by appropriate matching of pulse duration and crystal length. (A similar figure has been previously published in~\cite{laudenbach2016modelling} and was modified for this article.)}
\label{PvstauL}
\end{figure*}

Designing an experimental setup that generates photons of high intrinsic purity is only possible in specific configurations of wavelengths, polarisations and nonlinear crystal, where either $D \sim 0$ or $D \sim 1$. On top of that, once such a configuration is found, the length of the crystal has to be matched to the spectral bandwidth of the pump laser (especially for the case $D \sim 1$), as illustrated in \Cref{PvstauL}. A long crystal combined with a broadband laser will generate frequency-correlated daughter photons. Conversely, a short crystal and narrow-bandwidth laser will result in anti-correlated spectra. Both cases undermine the purity of the generated photon states.

\begin{figure*}
	\centering
\subcaptionbox{JSA}
	[0.49\linewidth]{\includegraphics[width=0.49\linewidth]{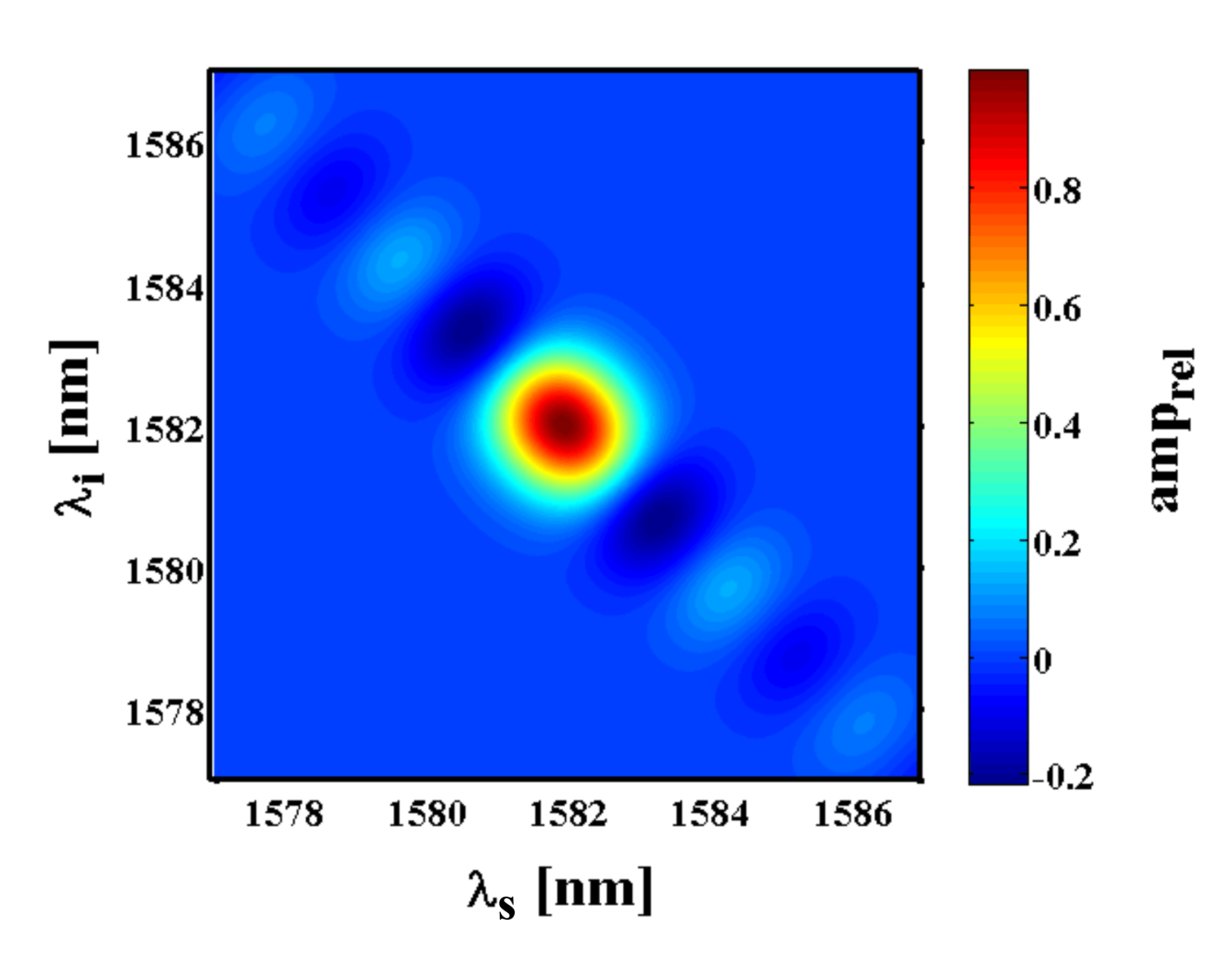}}
\subcaptionbox{JSI}
	[0.49\linewidth]{\includegraphics[width=0.49\linewidth]{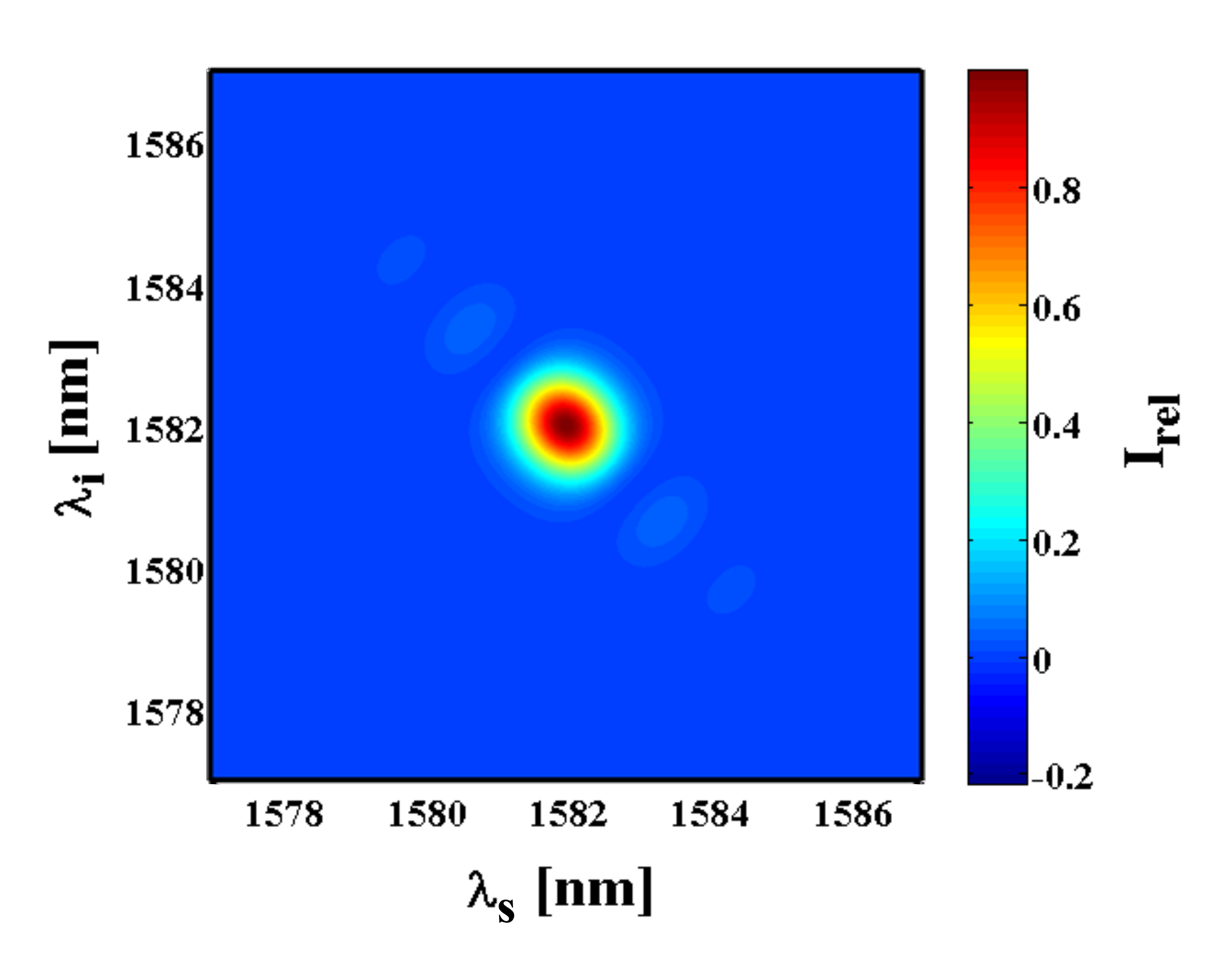}}
\subcaptionbox{JSA}
	[0.49\linewidth]{\includegraphics[width=0.49\linewidth]{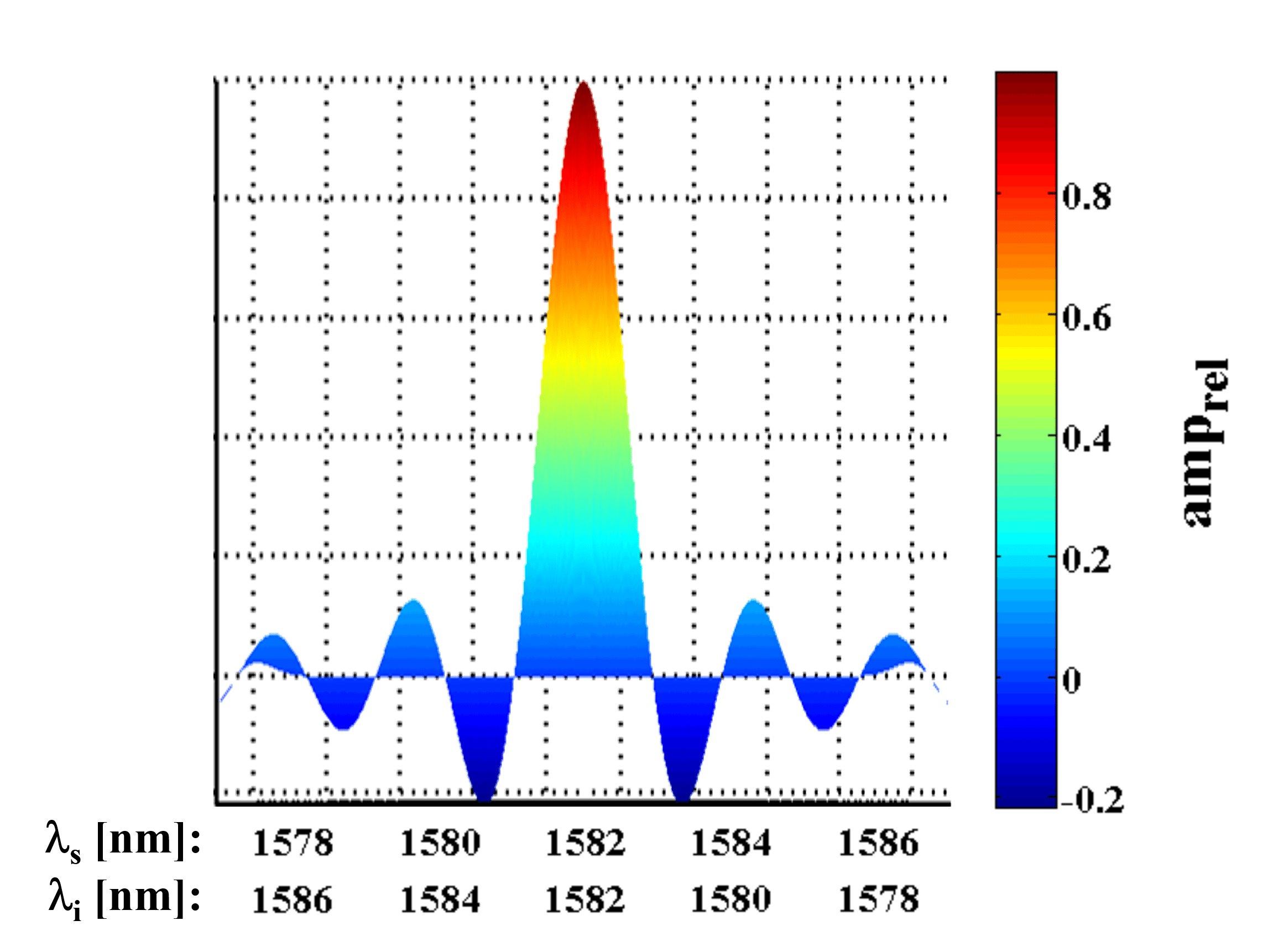}}
\subcaptionbox{JSI}
	[0.49\linewidth]{\includegraphics[width=0.49\linewidth]{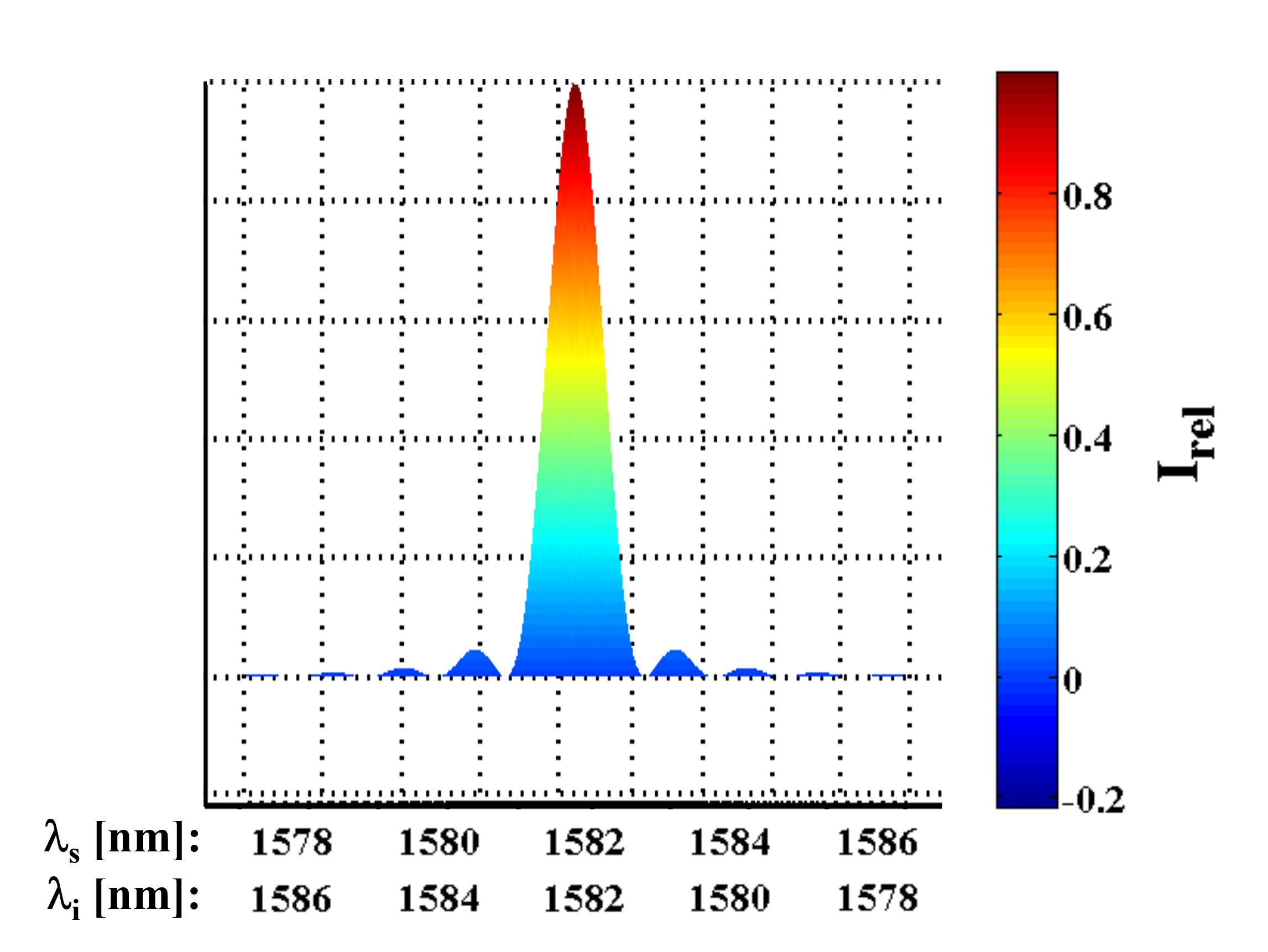}}
\caption{Spectral distribution of a frequency-degenerate type-II SPDC, $\SI{791}{\nano\metre}~(o) \longrightarrow \SI{1582}{\nano\metre}~(o) +\SI{1582}{\nano\metre}~(e)$, in a $\SI{30}{\milli\metre}$ ppKTP using a $\SI{2.5}{\pico\second}$ pump. Although the main peak is entirely frequency uncorrelated, the purity is still limited ($P\sim 0.83$, $K\sim 1.2$) due to the side lobes in the joint spectral amplitude (a). In the joint spectral intensity distribution (b), however, the side lobes are suppressed by squaring of the JSA, resulting in a very low Schmidt number $K_{\text{JSI}} ~\sim 1.01$. Figures~(c) and (d) depict a diagonal cross section of the 3D representation of (a) and (b), respectively.}
\label{JSAJSI}
\end{figure*}

\begin{figure*}
\centering
\includegraphics[width=.94\linewidth]{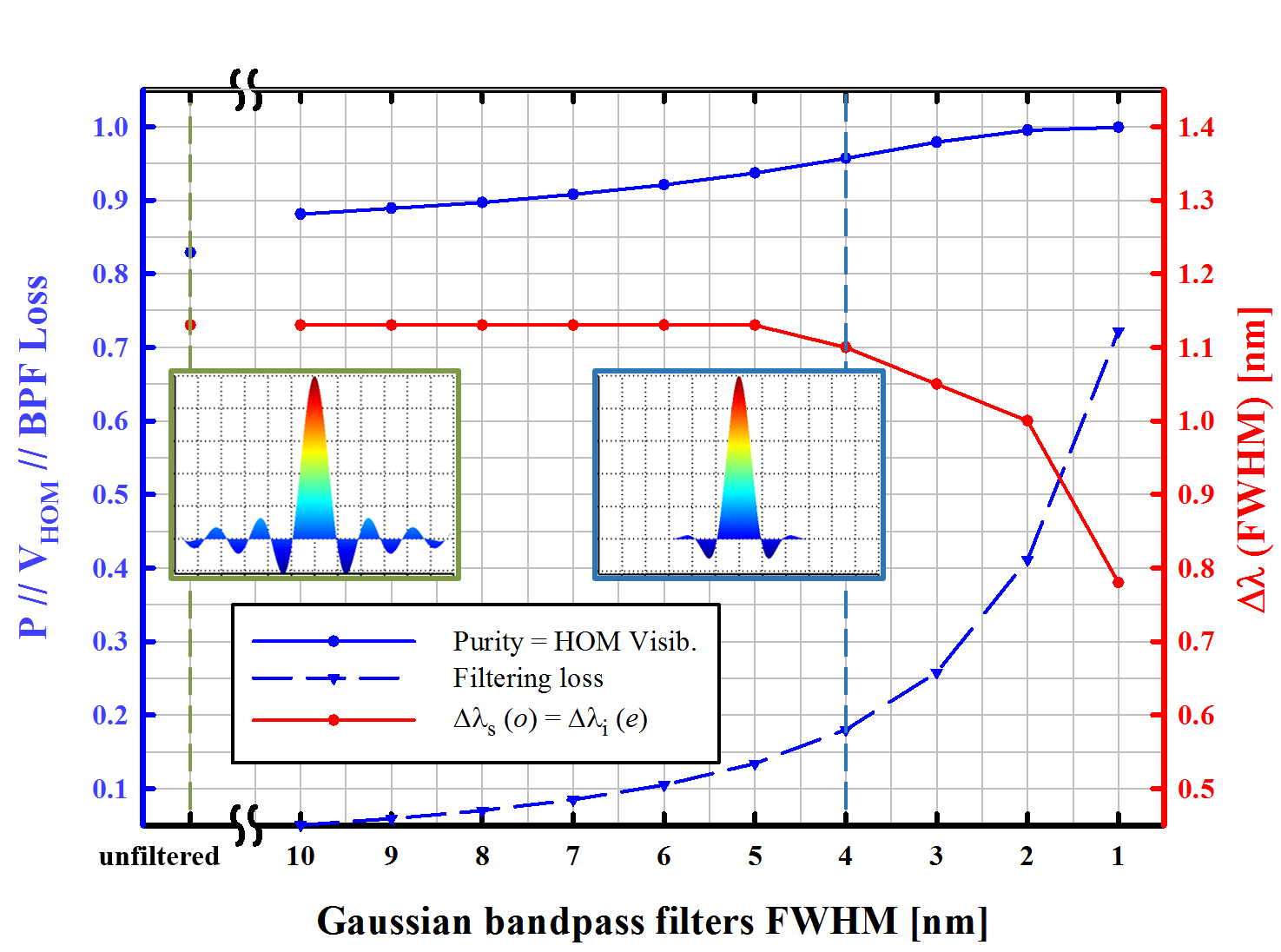}
\caption{The effects of bandpass filtering at frequency-degenerate type-II SPDC, $\SI{791}{\nano\metre}~(o) \longrightarrow \SI{1582}{\nano\metre}~(o) +\SI{1582}{\nano\metre}~(e)$, in a $\SI{30}{\milli\metre}$ ppKTP using a $\SI{2.5}{\pico\second}$ pump. The $x$-axis represents the full width at half maximum (FWHM) of Gaussian bandpass filters in the signal and idler channel. The solid blue line represents the spectral purity $P$ which coincides with the Hong-Ou-Mandel visibility $V_{\text{HOM}}$ since for this setup signal and idler are spectrally indistinguishable ($\lambda_{s}=\lambda_{i}$, $\Delta \lambda_{s}=\Delta  \lambda_{i}$, $\Delta=0$). The intensity losses due to filtering are illustrated by the dashed blue line. The red line indicates the FWHM of the filtered photons. The left inset depicts the JSA of the unfiltered case ($P\sim 0.83$); the right inset illustrates the JSA under $\SI{4}{\nano\metre}$-bandpass filtering. The vertical dashed sky-blue line indicates that the purity can be improved to $P>0.95$ under intensity losses of less than $20~\%$.}
\label{BPFKTPooe}
\end{figure*}

Even in cases where the experimentally observed spectral distribution of signal and idler is almost uncorrelated ($K_{\text{JSI}} \leq 1.01$), interference visibilities might still be undermined due to correlations in the JSA~\cite{Gerrits2014}. These spectral correlations are introduced by side lobes in the JSA, caused by the sinc function in the QPM amplitude $\psi$. These side lobes are suppressed in the JSI by squaring of the JSA, as depicted in \Cref{JSAJSI}. Although the actual HOM visibility depends on the factorability of the JSA, it nonetheless makes sense to pay attention to $K_{\text{JSI}}$ as well, since a low Schmidt number of $K_{\text{JSI}} \leq 1.01$ indicates that the main peak of the joint spectral distribution is factorable and the purity is undermined by the side lobes only. These side lobes, however, can be removed by bandpass filtering under relatively low intensity loss, as illustrated by an example in \Cref{BPFKTPooe}. In this particular example identical Gaussian-shape bandpass filters with ideal transmission (100~\%) on each photon are considered for a particular QPM process, i.e.\ $\SI{791}{\nano\metre} \ (o) \longrightarrow \SI{1582}{\nano\metre} \ (o) + \SI{1582}{\nano\metre} \ (e)$. As shown in the plot, the purity extracted from the HOM visibility between independent photons increases by cutting the side lobes of the joint spectrum through the filters while introducing a reasonable amount of loss and therefore maintaining high count rates. While for this example we assumed identical filters in each of the four arms of two SPDC sources, we have shown previously that already two out of four possible filters are able to enhance the purity substantially~\cite{Jin2015swapping}. In the remainder of this article, we speak of `intrinsically pure' SPDC when $K_{\text{JSI}} \leq 1.01$, hence when high purity can be achieved without or with very low-loss filtering.

Note that the simulations in this article do not include a study of the spatial correlations between the downconverted photons. These become relevant when the photons are collected via single mode fibres in order to be processed and detected. Taking the spatial correlations into account \cite{Guerriero2013, Grice2011, Bruno2014, Kaneda2016, Gajewski2016}, it is possible to cut the side lobes of the JSA directly through spatial filtering of the single-mode coupling: Assuming the pump beam has a large waist at the centre of the crystal, one can find the specific configuration of single mode fibres and coupling lenses for which the spatial correlations increase and the frequency correlations reduce. Increasing the beam waist decreases the total coupling efficiency and therefore, in a similar way to the bandpass-filtering approach, one should find the optimal design for high purity and count rates as required by the experiment.

This section provides a collection and discussion of configurations in which uncorrelated spectra can in principle be achieved\textemdash provided matched laser bandwidth and crystal length. The calculations are based on the dispersion equations and other material properties of KTP, CTA, KTA, RTA and RTP \cite{Konig2004, Fradkin1999, Kato2002, Mikami2009, Kato1994, Cheng1994, Cheng1993}. Note that only type-0 and type-II downconversion is mentioned since, according to our calculations, KTP and its isomorphs do not allow for intrinsically pure states at type-I SPDC (signal and idler parallelly polarised, but orthogonally with respect to the pump). However, pure states generated by type-I downconversion are possible using periodically poled lithium niobate and lithium tantalate~\cite{laudenbach2016modelling}.

\subsection{Type 0}

\begin{figure*}
\centering
\includegraphics[width=.94\linewidth]{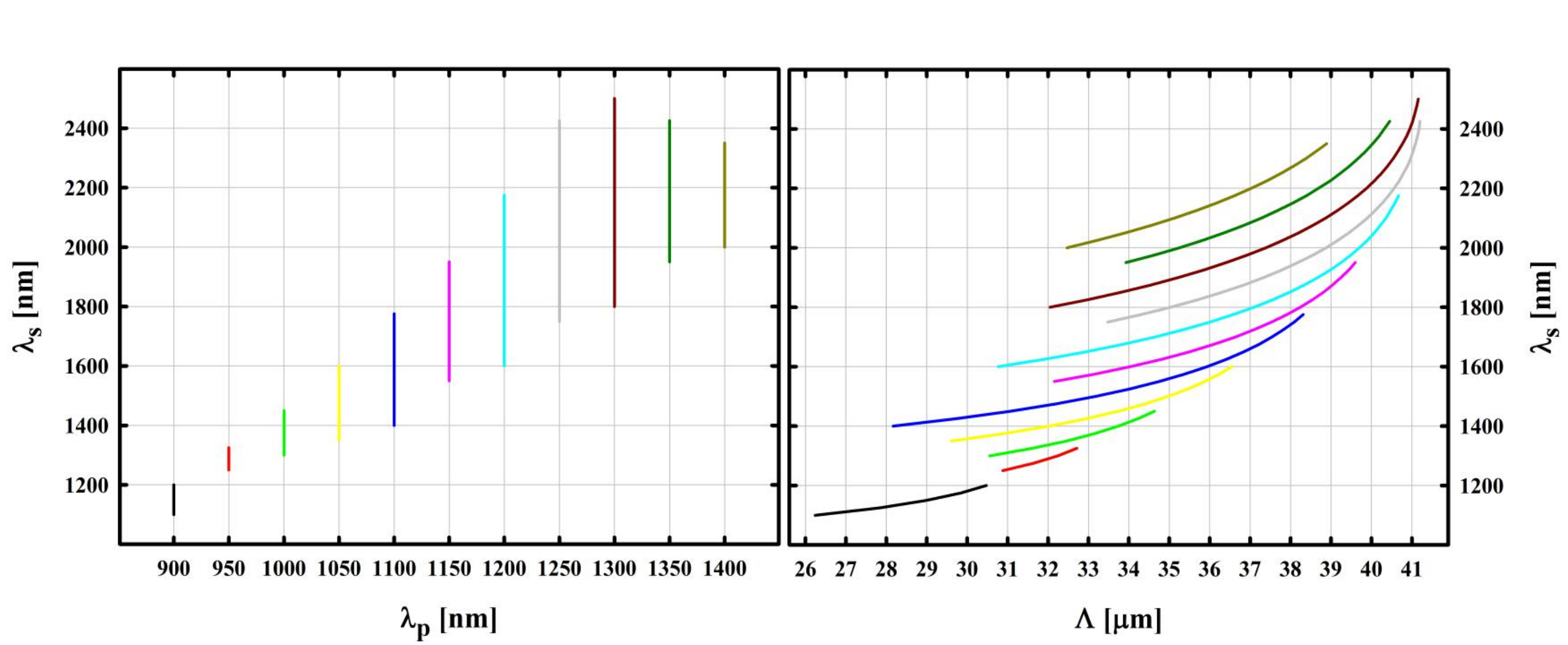}
\caption{Signal wavelength versus pump wavelength $\lambda_{p}$ (left) and crystal periodicity $\Lambda$ (right) for pure type-0 SPDC ($e \longrightarrow e + e$) in ppKTP ($d_{\text{eff}} \sim \SI{9.5}{\pico\metre\volt^{-1}}$). The plots display configurations which allow for spectrally uncorrelated signal- and idler-photon states under no or low-loss bandpass filtering. Each coloured line corresponds to a particular pump wavelength $\lambda_{p}$ and depicts the range of signal wavelength over which a Schmidt number of $K_{\text{JSI}} \leq 1.01$ can be achieved by mutual matching of pump spectrum and crystal length. Note that all depicted data corresponds to a crystal temperature of \SI{50}{\degreeCelsius} and is slightly modified with varying temperature. (This figure has been previously published in~\cite{laudenbach2016modelling}.)}
\label{KTP_eee}
\end{figure*}

\begin{figure*}
\centering
\includegraphics[width=.94\linewidth]{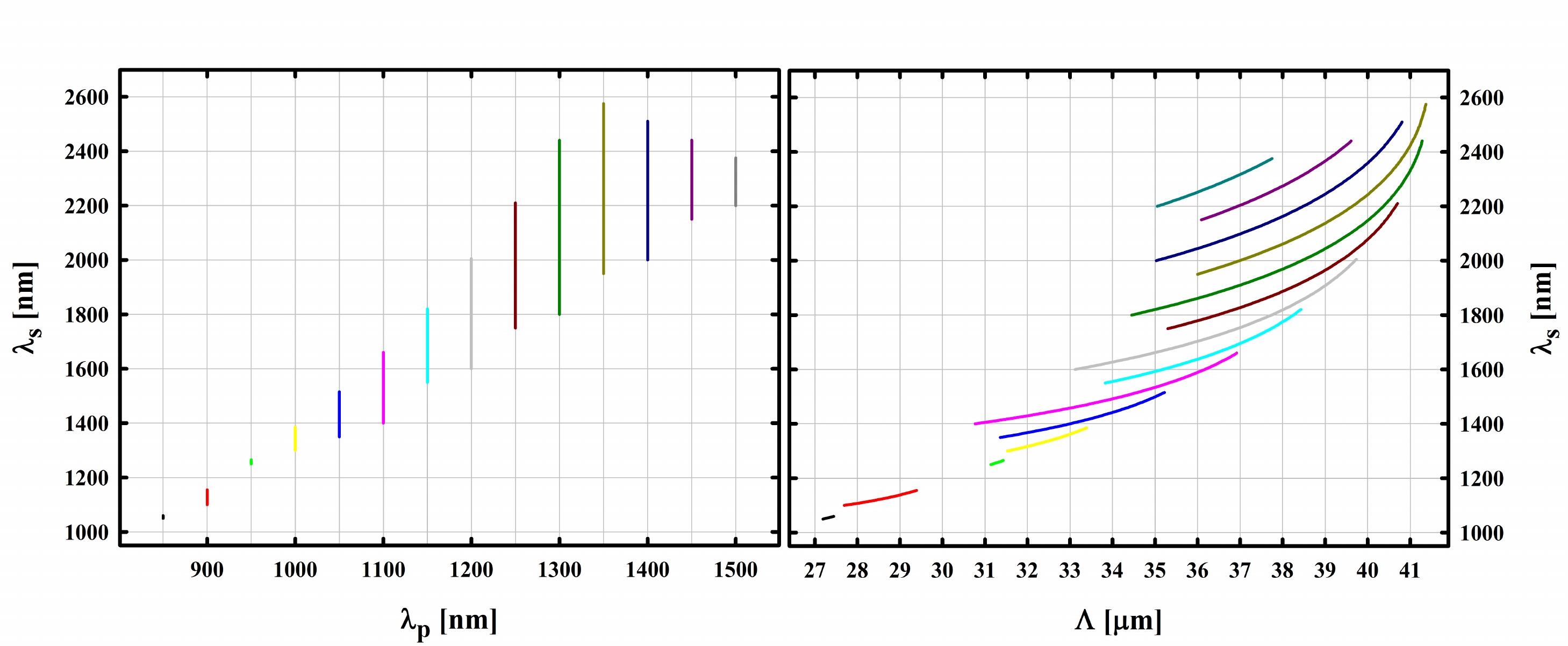}
\caption{Signal wavelength versus pump wavelength $\lambda_{p}$ (left) and crystal periodicity $\Lambda$ (right) for pure type-0 SPDC ($e \longrightarrow e + e$) in ppCTA ($d_{\text{eff}} \sim \SI{11.2}{\pico\metre\volt^{-1}}$)}
\label{CTA_eee}
\end{figure*}

\begin{figure*}
\centering
\includegraphics[width=.94\linewidth]{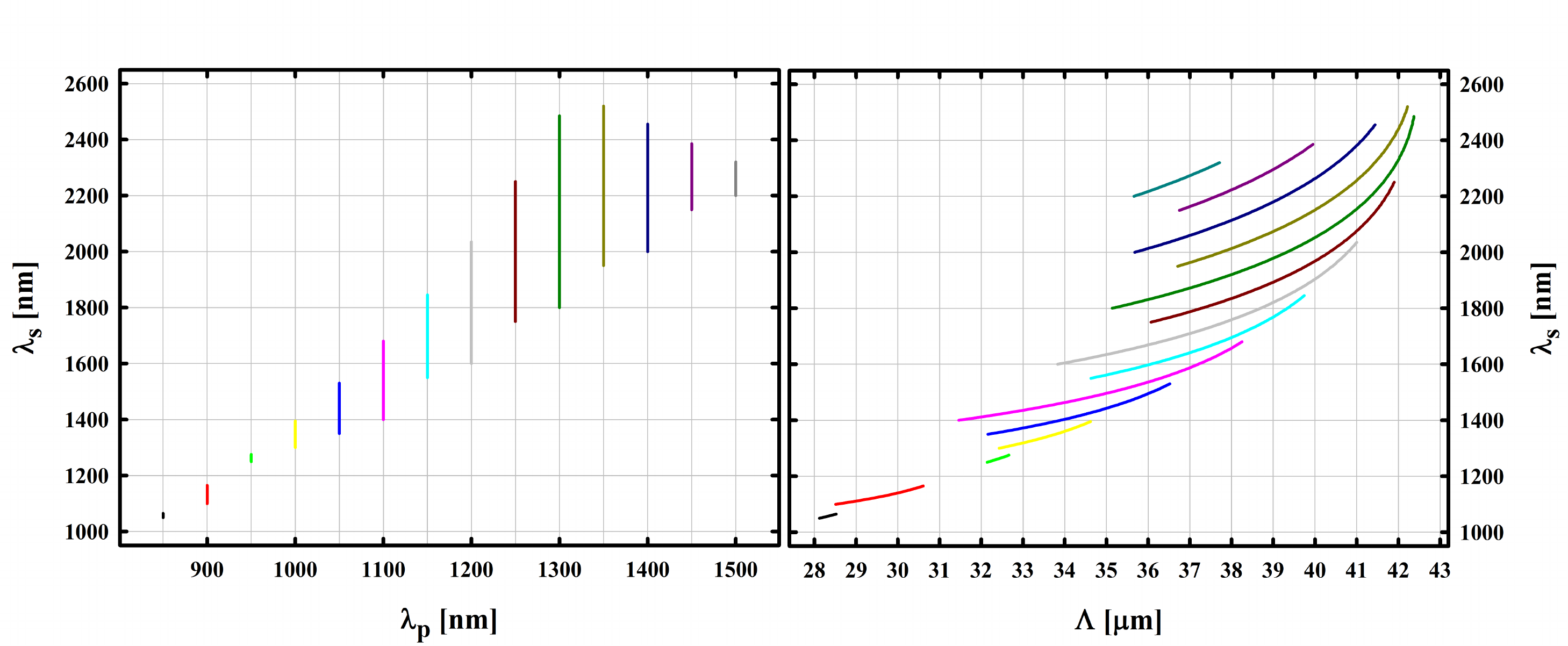}
\caption{Signal wavelength versus pump wavelength $\lambda_{p}$ (left) and crystal periodicity $\Lambda$ (right) for pure type-0 SPDC ($e \longrightarrow e + e$) in ppKTA ($d_{\text{eff}} \sim \SI{9.6}{\pico\metre\volt^{-1}}$)}
\label{KTA_eee}
\end{figure*}

\begin{figure*}
\centering
\includegraphics[width=.94\linewidth]{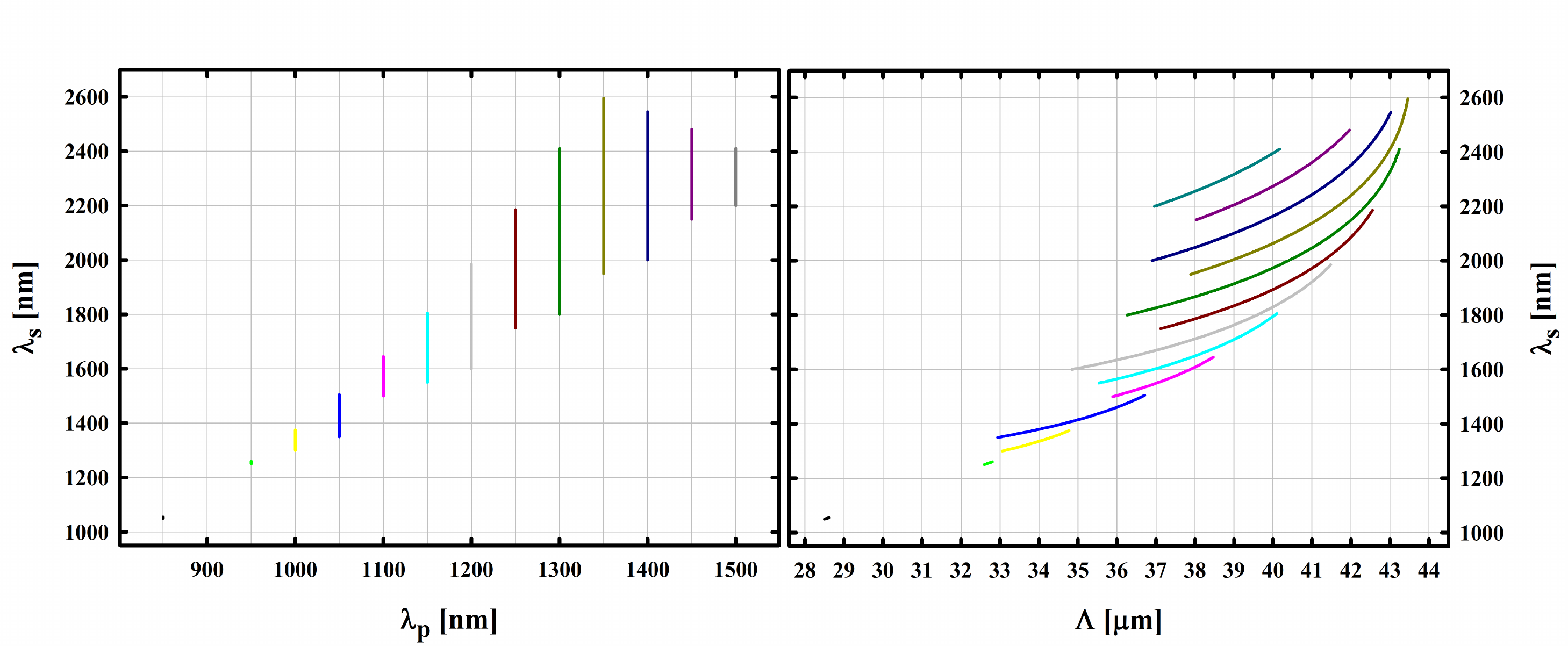}
\caption{Signal wavelength versus pump wavelength $\lambda_{p}$ (left) and crystal periodicity $\Lambda$ (right) for pure type-0 SPDC ($e \longrightarrow e + e$) in ppRTA ($d_{\text{eff}} \sim \SI{9.8}{\pico\metre\volt^{-1}}$)}
\label{RTA_eee}
\end{figure*}

\begin{figure*}
\centering
\includegraphics[width=.94\linewidth]{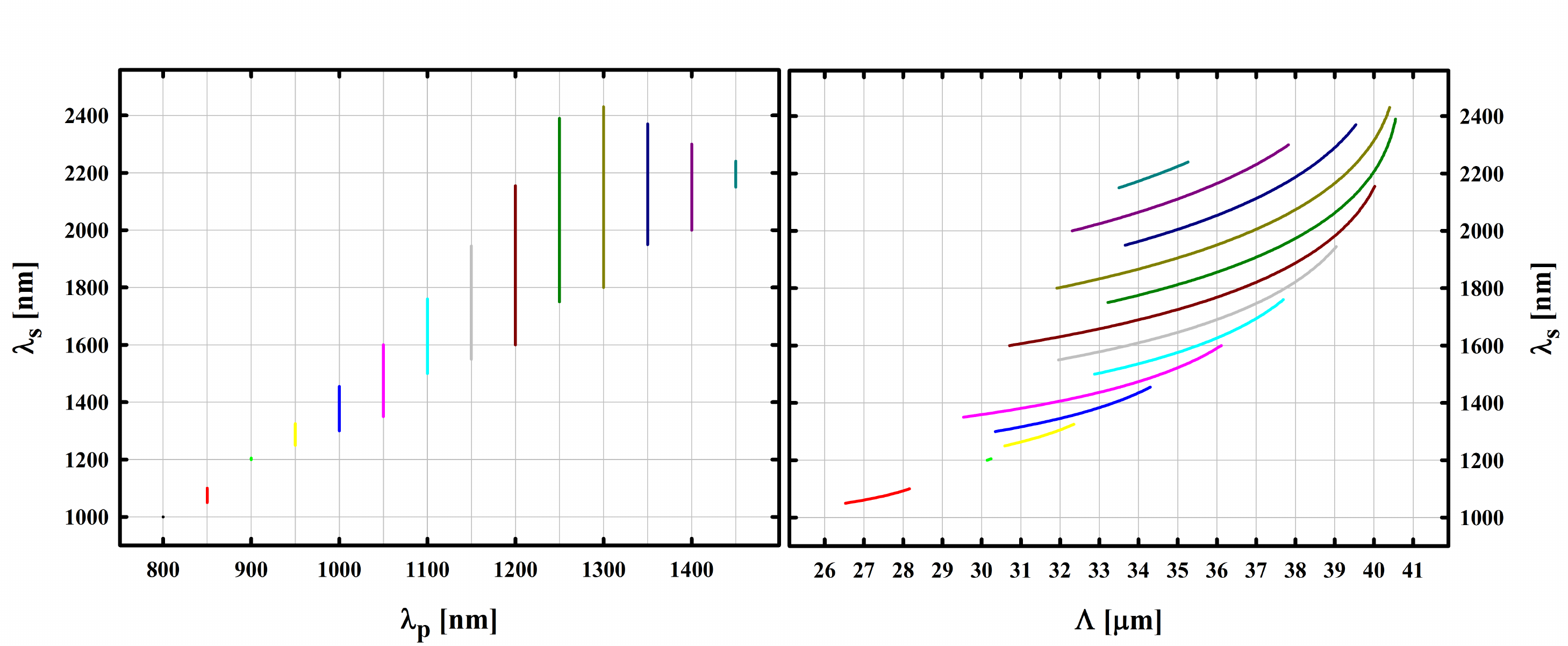}
\caption{Signal wavelength versus pump wavelength $\lambda_{p}$ (left) and crystal periodicity $\Lambda$ (right) for pure type-0 SPDC ($e \longrightarrow e + e$) in ppRTP  ($d_{\text{eff}} \sim \SI{9.6}{\pico\metre\volt^{-1}}$)}
\label{RTP_eee}
\end{figure*}

When signal and idler are not required to be orthogonally polarised (i.e.\ in non-degenerate SPDC processes), type-0 downconversion is often preferred over other polarisation configurations due to higher effective nonlinearities and hence greater pair-generation rates. \Cref{KTP_eee,CTA_eee,KTA_eee,RTA_eee,RTP_eee} describe type-0 SPDC processes which allow for high intrinsic purity when the crystal length and pump pulse duration are mutually matched. The plots on the left hand side depict pump- and signal-wavelength combinations for which a Schmidt number $K_{\text{JSI}}$ smaller than or equal 1.01 is possible. The plots on the right hand side illustrate the very same states (each colour representing a particular pump wavelength), but with respect to the poling period $\Lambda$ of the nonlinear crystal. Unfortunately, as the plots illustrate, high intrinsic purity at type-0 SPDC is only allowed for in rather unfavourable wavelength configurations where the signal or/and idler photon are too long-waved to be processed and detected efficiently.

\subsection{Type II}

\begin{figure*}
\centering
\includegraphics[width=.94\linewidth]{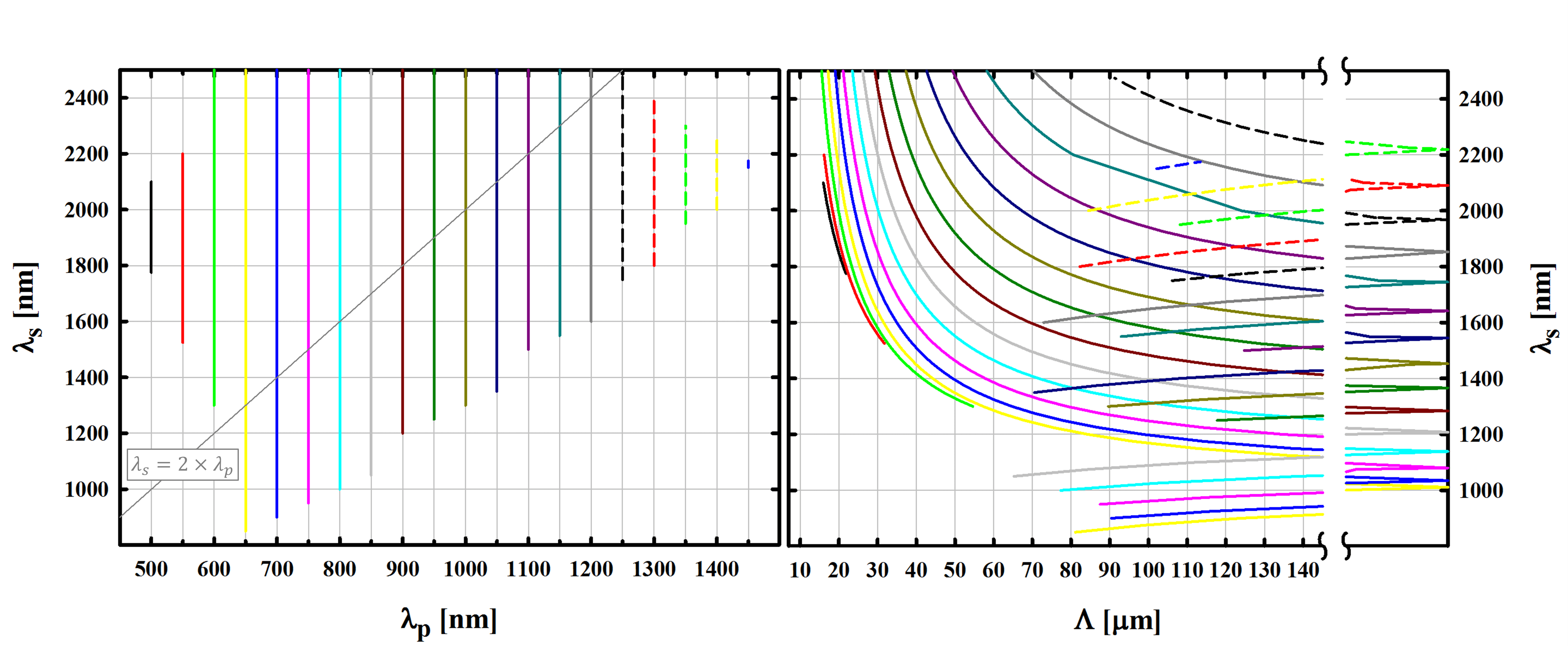}
\caption{Signal wavelength versus pump wavelength $\lambda_{p}$ (left) and crystal periodicity $\Lambda$ (right) for high intrinsic purity at type-II SPDC ($o \longrightarrow o + e$) in ppKTP ($d_{\text{eff}}=d^{32} \sim \SI{2.4}{\pico\metre\volt^{-1}}$) at $T=\SI{50}{\degreeCelsius}$. The grey line in the left plot represents spectrally symmetric downconversions. The right plot indicates a range of pump and signal wavelengths for which the periodicity $\Lambda$ approaches infinity, thus enabling spectrally pure output generation without periodic poling (displayed in \Cref{CTAnopoling,KTPKTARTARTPnopoling}). (A similar figure has been previously published in~\cite{laudenbach2016modelling} and was modified for this article.)}
\label{KTP_ooe}
\end{figure*}

\begin{figure*}
\centering
\includegraphics[width=.94\linewidth]{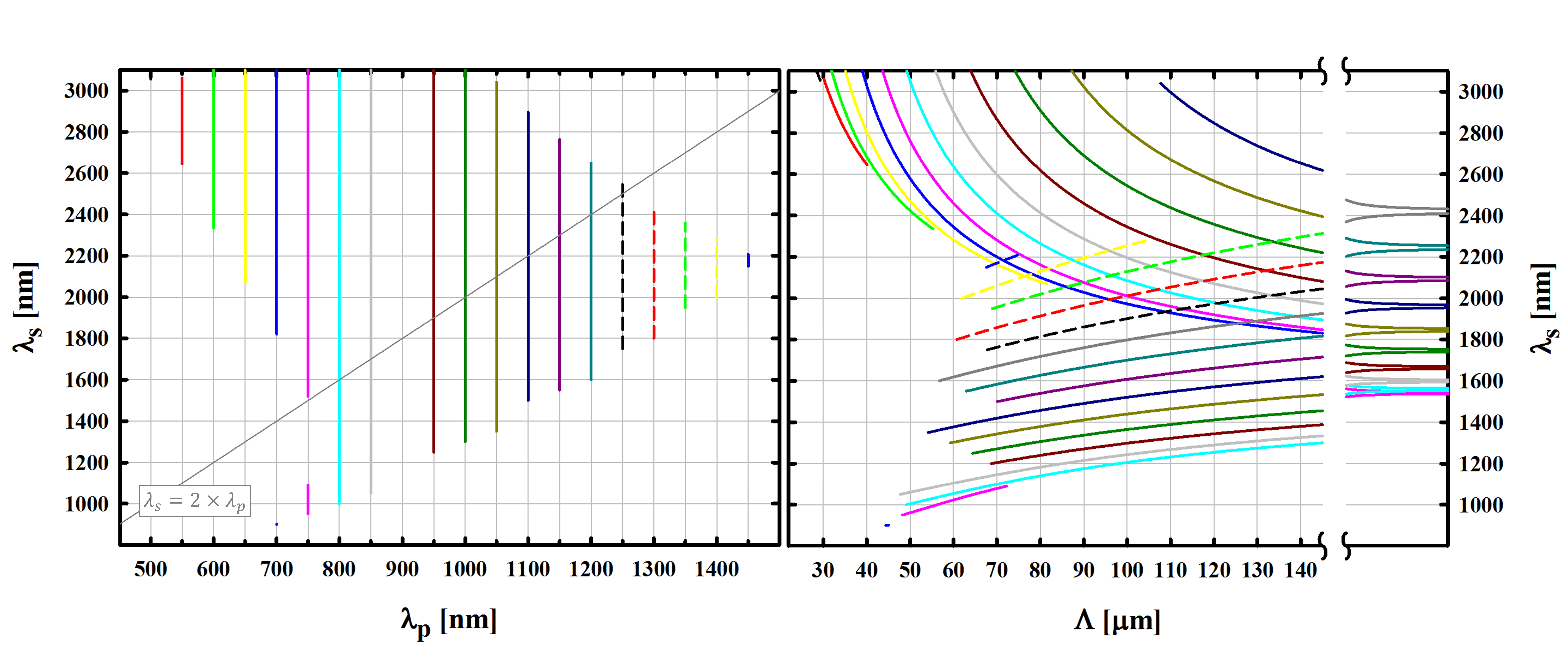}
\caption{Signal wavelength versus pump wavelength $\lambda_{p}$ (left) and crystal periodicity $\Lambda$ (right) for high intrinsic purity at type-II SPDC ($o \longrightarrow o + e$) in ppCTA ($d_{\text{eff}}=d^{32} \sim \SI{2.1}{\pico\metre\volt^{-1}}$)}
\label{CTA_ooe}
\end{figure*}

\begin{figure*}
\centering
\includegraphics[width=.94\linewidth]{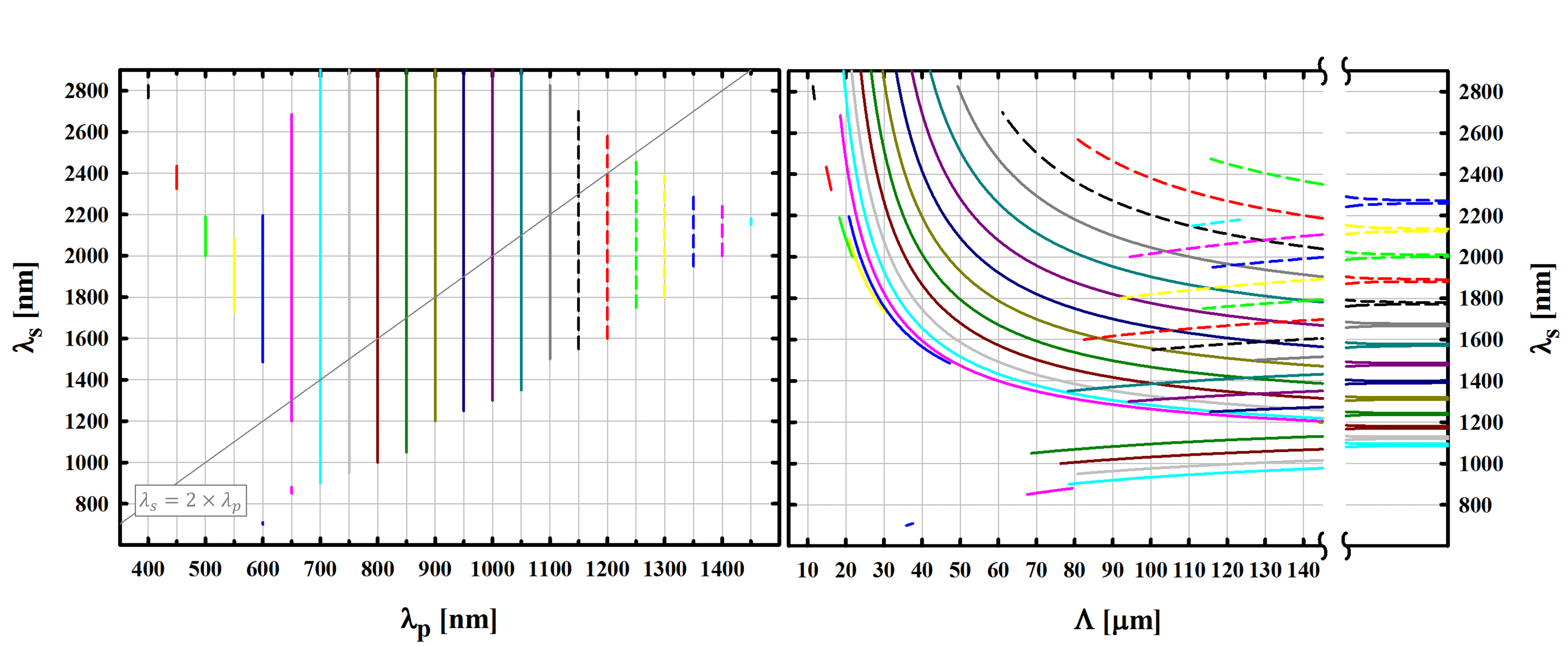}
\caption{Signal wavelength versus pump wavelength $\lambda_{p}$ (left) and crystal periodicity $\Lambda$ (right) for high intrinsic purity at type-II SPDC ($o \longrightarrow o + e$) in ppKTA ($d_{\text{eff}}=d^{32} \sim \SI{2.3}{\pico\metre\volt^{-1}}$)}
\label{KTA_ooe}
\end{figure*}

\begin{figure*}
\centering
\includegraphics[width=.94\linewidth]{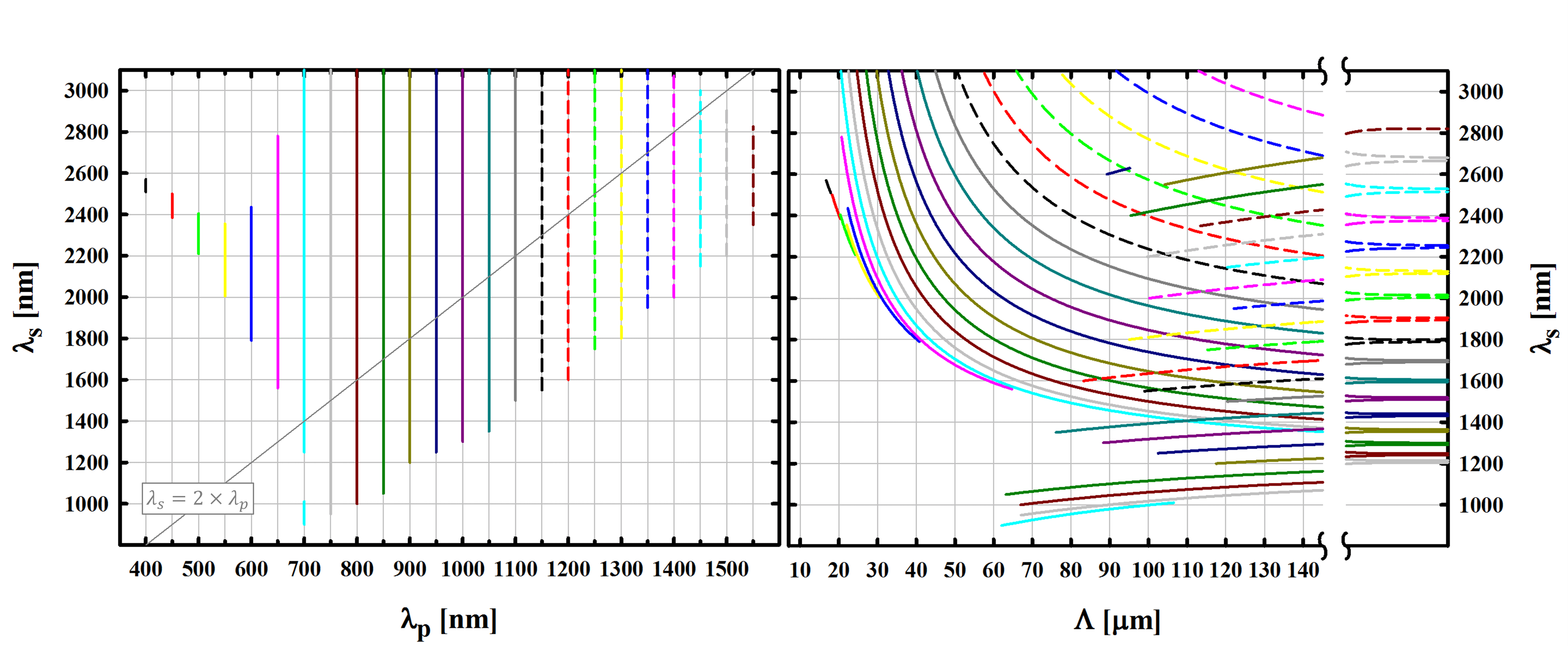}
\caption{Signal wavelength versus pump wavelength $\lambda_{p}$ (left) and crystal periodicity $\Lambda$ (right) for high intrinsic purity at type-II SPDC ($o \longrightarrow o + e$) in ppRTA ($d_{\text{eff}}=d^{32} \sim \SI{2.4}{\pico\metre\volt^{-1}}$)}
\label{RTA_ooe}
\end{figure*}

\begin{figure*}
\centering
\includegraphics[width=.94\linewidth]{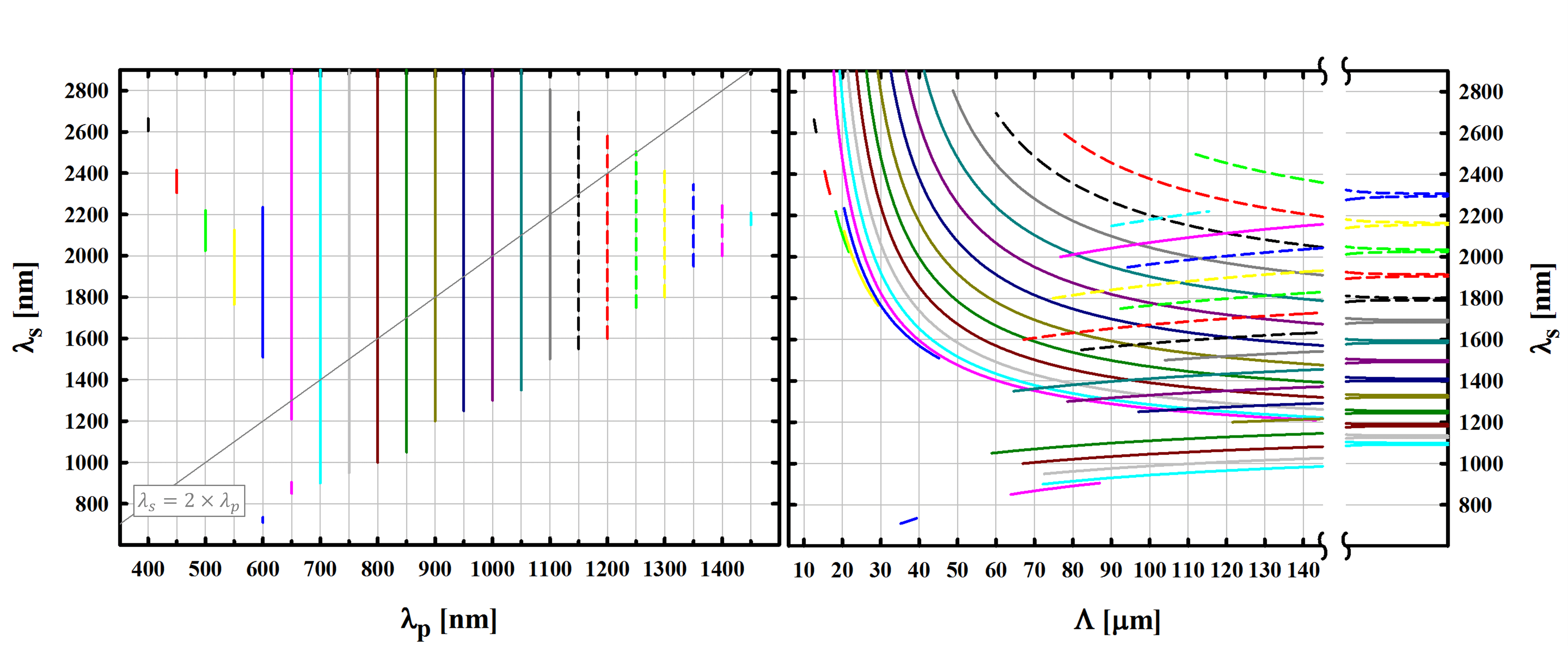}
\caption{Signal wavelength versus pump wavelength $\lambda_{p}$ (left) and crystal periodicity $\Lambda$ (right) for high intrinsic purity at type-II SPDC ($o \longrightarrow o + e$) in ppRTP ($d_{\text{eff}}=d^{32} \sim \SI{2.4}{\pico\metre\volt^{-1}}$)}
\label{RTP_ooe}
\end{figure*}

Many experiments require photon pairs of orthogonal polarisation, allowing e.g.\ for an easy separation of the pairs into different spatial modes and generation of polarisation entanglement. Find in this section a collection of intrinsically pure type-II downconversion states, depicted in \Cref{KTP_ooe,CTA_ooe,KTA_ooe,RTA_ooe,RTP_ooe}. The grey line on the left graphs represents the important special case of frequency-degenerate downconversion. These configurations are more deeply investigated, in terms of purity and indistinguishability, in Section~\ref{sec_degenerate}. The plots on the right side indicate that for all five materials there is a set of wavelength configurations for which the required poling period approaches infinity. This interesting feature allows for collinear and intrinsically pure SPDC in \emph{bulk} crystals with no periodic poling required (closer investigated in Section~\ref{sec_nopoling}).

\subsubsection{Frequency-degenerate SPDC} \label{sec_degenerate}

\begin{figure*}
\centering
\includegraphics[width=0.77\linewidth]{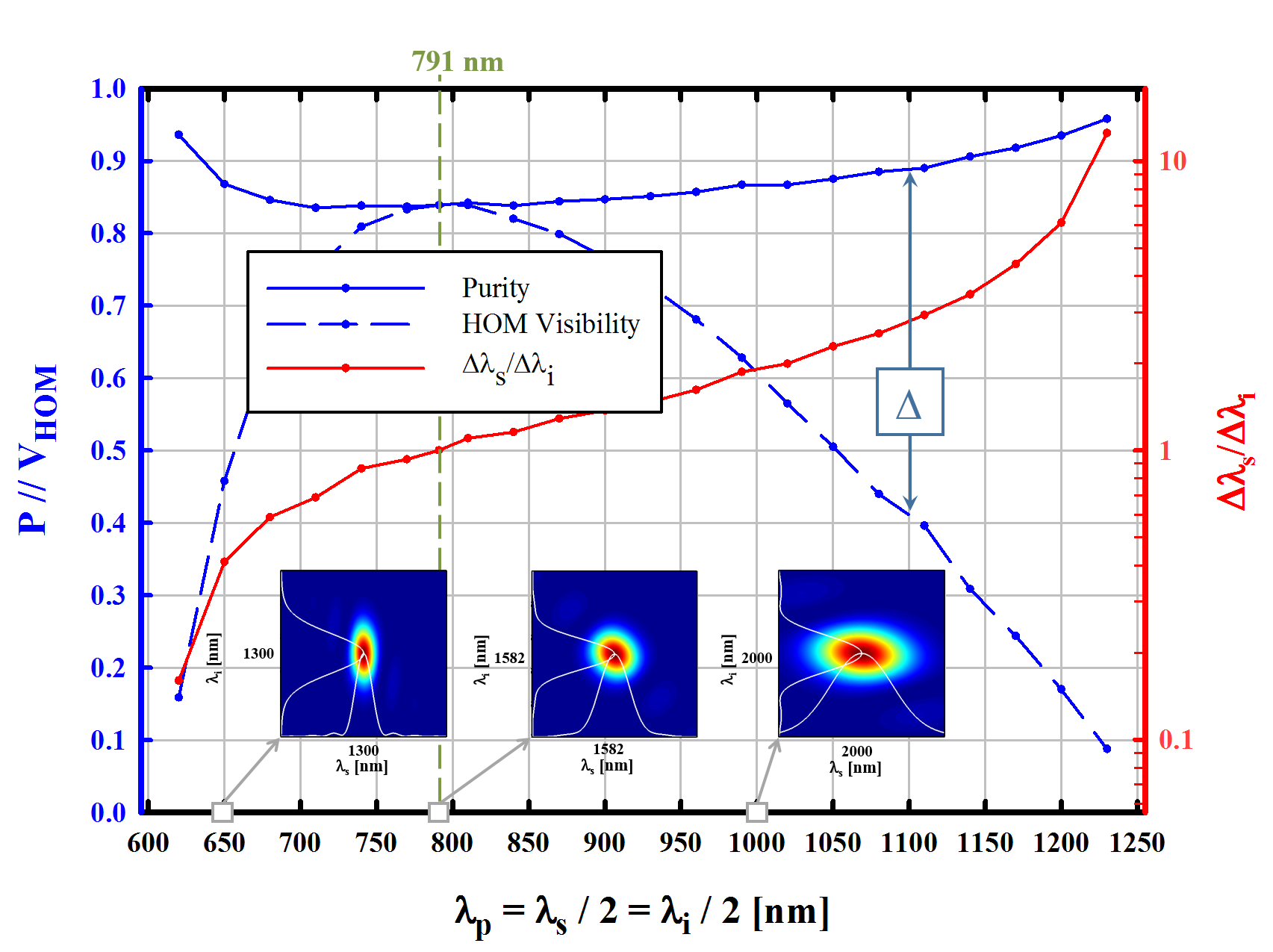}
\caption{Intrinsically pure frequency-degenerate type-II SPDC states with ppKTP. For each pump wavelength on the $x$-axis the crystal length and pump bandwidth have been mutually matched to maximise the purity. The solid blue line represents the maximally achievable spectral purity $P$ whereas the dashed blue line depicts the HOM visibility $V_{\text{HOM}}=P-\Delta$ where $\Delta$ is a measure for spectral distinguishability. The red line represents the ratio of signal and idler bandwidth. Maximal spectral indistinguishability can be achieved at $\lambda_{p}=\SI{791}{\nano\metre}$, as indicated by $\Delta \lambda_{s} / \Delta \lambda_{i} = 1$ and the intersection of the blue lines ($P=V_{\text{HOM}}$). (A similar figure has been previously published in~\cite{laudenbach2016modelling} and was modified for this article.)}
\label{KTPooe_HOM}
\end{figure*}

\begin{figure*}
	\centering
\subcaptionbox{ppCTA, $\Delta=0$ at $\lambda_{p}=\SI{932.3}{\nano\metre}$}
	[0.49\linewidth]{\includegraphics[width=0.49\linewidth]{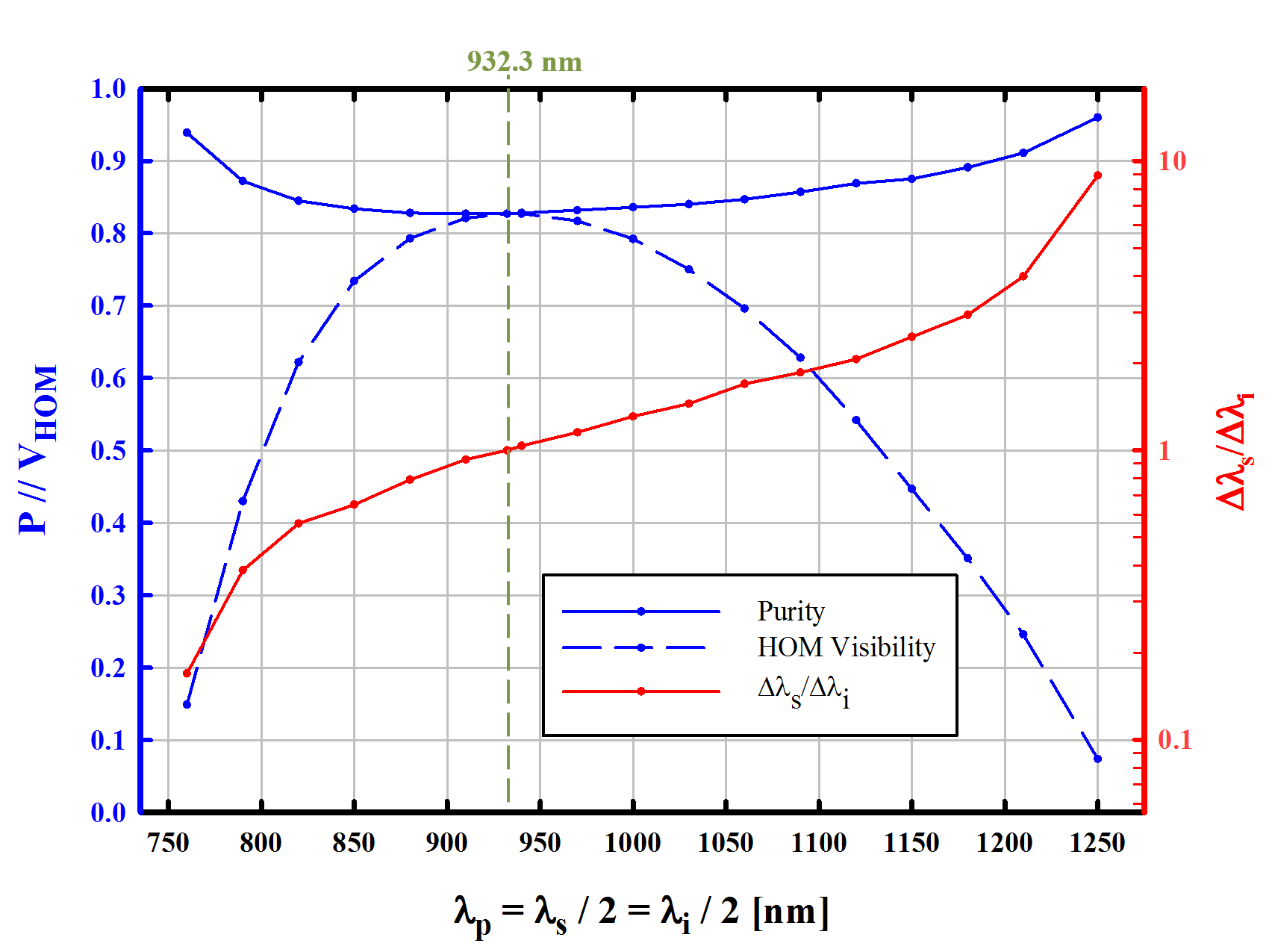}}
\subcaptionbox{ppKTA, $\Delta=0$ at $\lambda_{p}=\SI{817.4}{\nano\metre}$}
	[0.49\linewidth]{\includegraphics[width=0.49\linewidth]{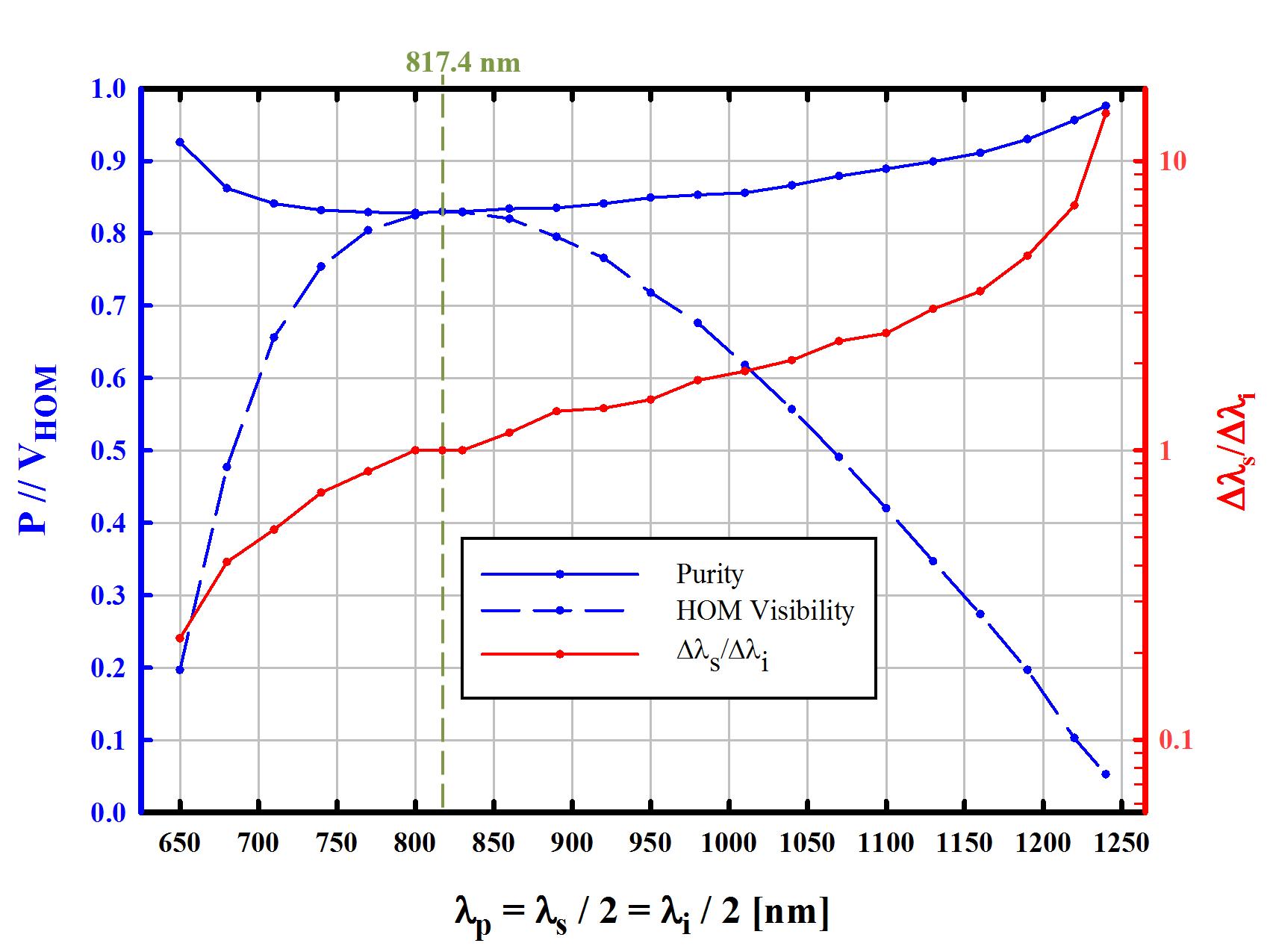}}
\subcaptionbox{ppRTA, $\Delta=0$ at $\lambda_{p}=\SI{892.3}{\nano\metre}$}
	[0.49\linewidth]{\includegraphics[width=0.49\linewidth]{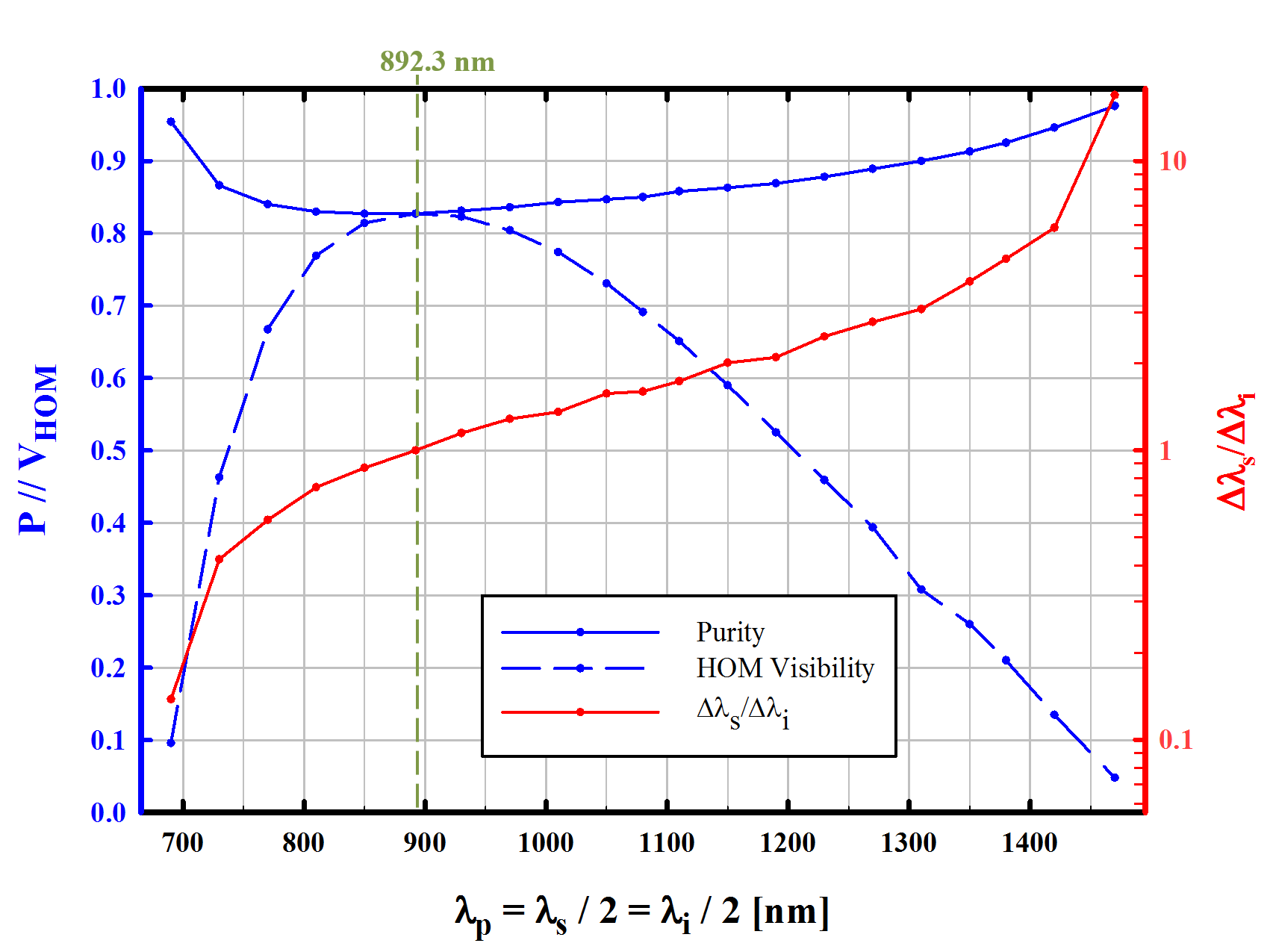}}
\subcaptionbox{ppRTP, $\Delta=0$ at $\lambda_{p}=\SI{821.6}{\nano\metre}$}
	[0.49\linewidth]{\includegraphics[width=0.49\linewidth]{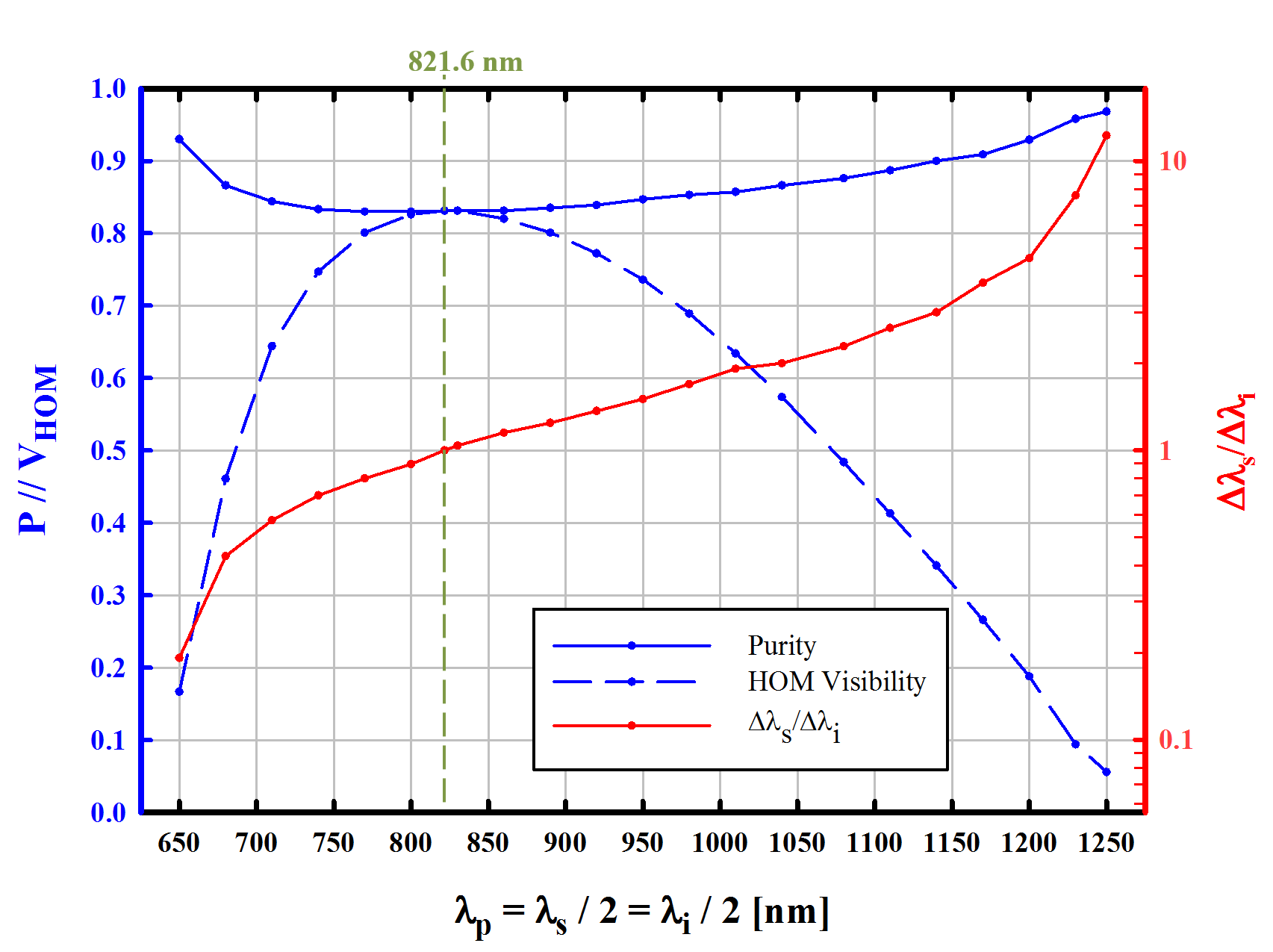}}
\caption{Intrinsically pure frequency-degenerate type-II SPDC with (a)~\text{ppCTA}, (b)~\text{ppKTA}, (c)~\text{ppRTA} and (d)~\text{ppRTP}.}
\label{CTAKTARTARTPooe_HOM}
\end{figure*}

Photon pairs with the same centre wavelength but orthogonal polarisation find many applications in quantum optics. As illustrated in the previous section, each crystal comes with a large set of possible spectrally pure and frequency-degenerate SPDC states. However, there is for each material only one point where the exact same centre wavelength \emph{and} spectral bandwidth can be achieved, hence where the dispersion parameter $D$ \eqref{dispersionparameter} equals one. These configurations of maximal spectral indistinguishability can be regarded as material constants of the respective nonlinear crystals. \Cref{KTPooe_HOM,CTAKTARTARTPooe_HOM} depict the maximally achievable purity and HOM visibility for frequency-degenerate downconversion with respect to the pump wavelength. The graphs illustrate that minimal indistinguishability $\Delta$ does not correspond to maximal spectral purity $P$. This is due to the spectral correlations carried by the side lobes of the sinc function in the QPM amplitude.

\subsubsection{Pure States w/o Periodic poling} \label{sec_nopoling}

\begin{figure*}
	\centering
	\includegraphics[width=0.77\linewidth]{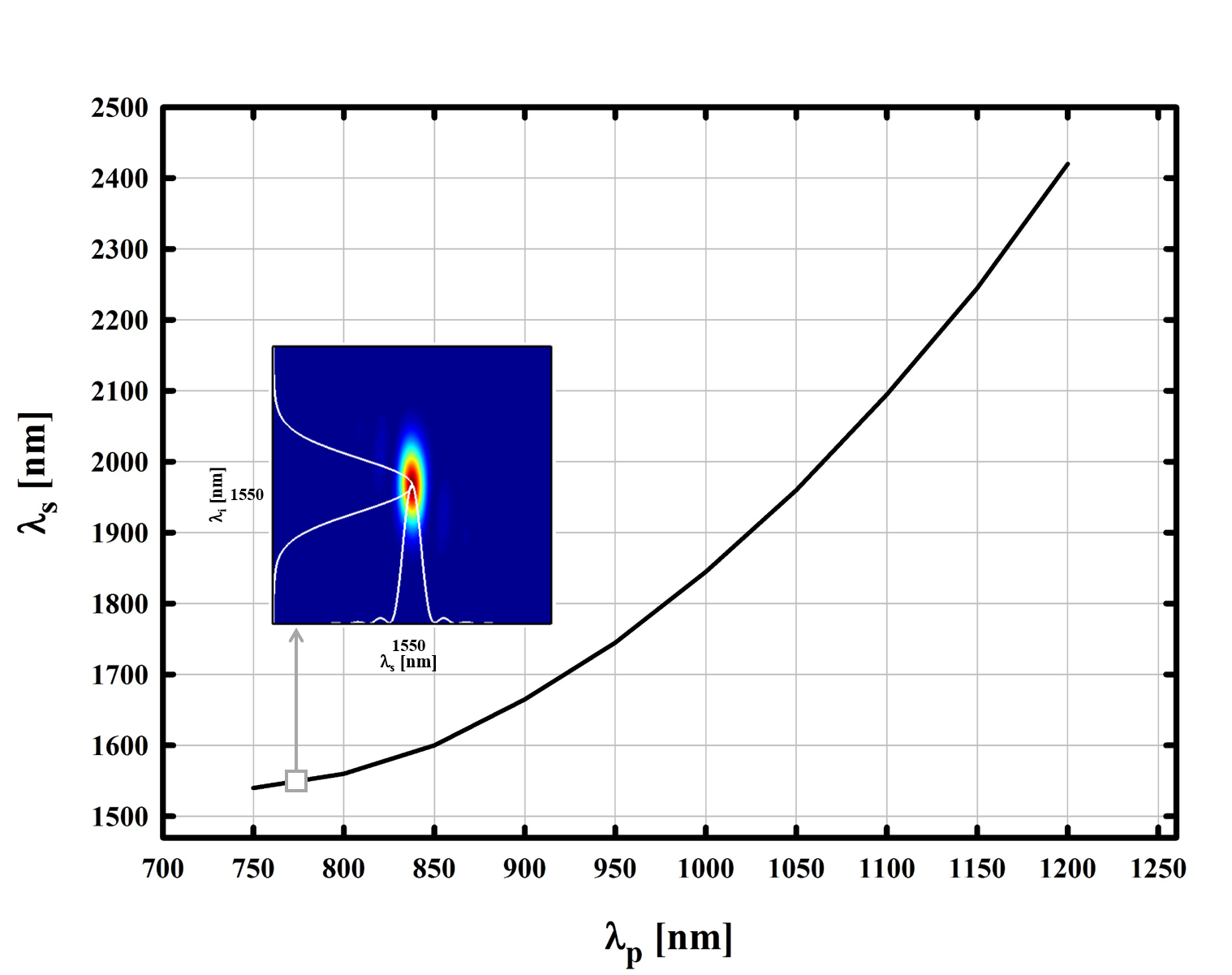}
\caption{Pure states with bulk CTA, $o \longrightarrow o + e$. Illustration of setups for type-II SPDC which allow for phase-matched and intrinsically pure states in bulk\textemdash not periodically poled\textemdash CTA. The idler wavelength corresponding to a pair of $\lambda_{p}$ and $\lambda_{s}$ is obtained by energy conservation. The inset corresponds to a collinear downconversion from $\SI{775}{\nano\metre}$ to two times $\SI{1550}{\nano\metre}$ in a CTA crystal without periodic ferroelectric poling and with a purity of $P\sim 0.91$.}
\label{CTAnopoling}
\end{figure*}

\begin{figure*}
	\centering
\subcaptionbox{KTP}
	[0.49\linewidth]{\includegraphics[width=0.525\linewidth]{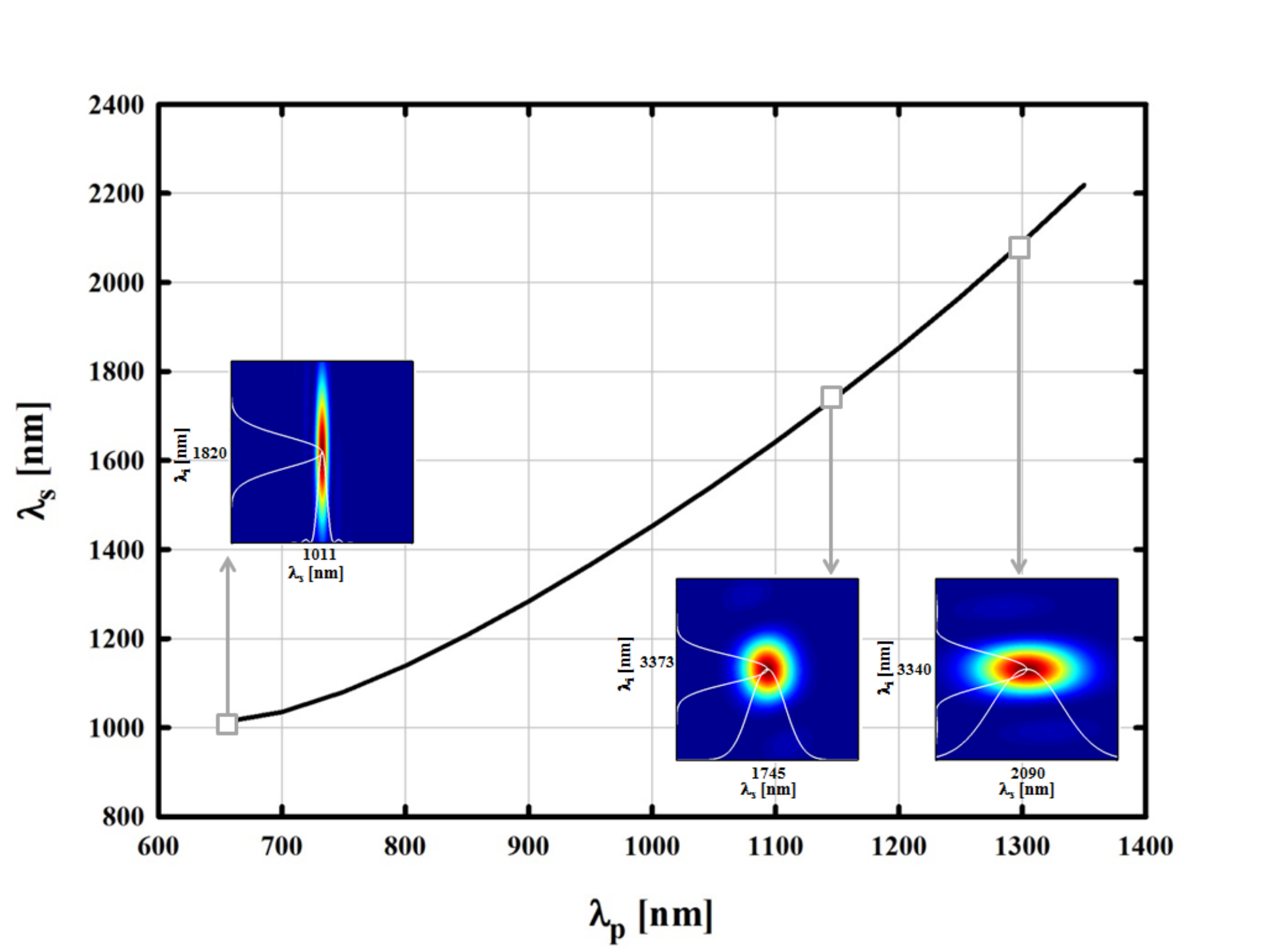}}
\subcaptionbox{KTA, RTA, RTP}
	[0.49\linewidth]{\includegraphics[width=0.49\linewidth]{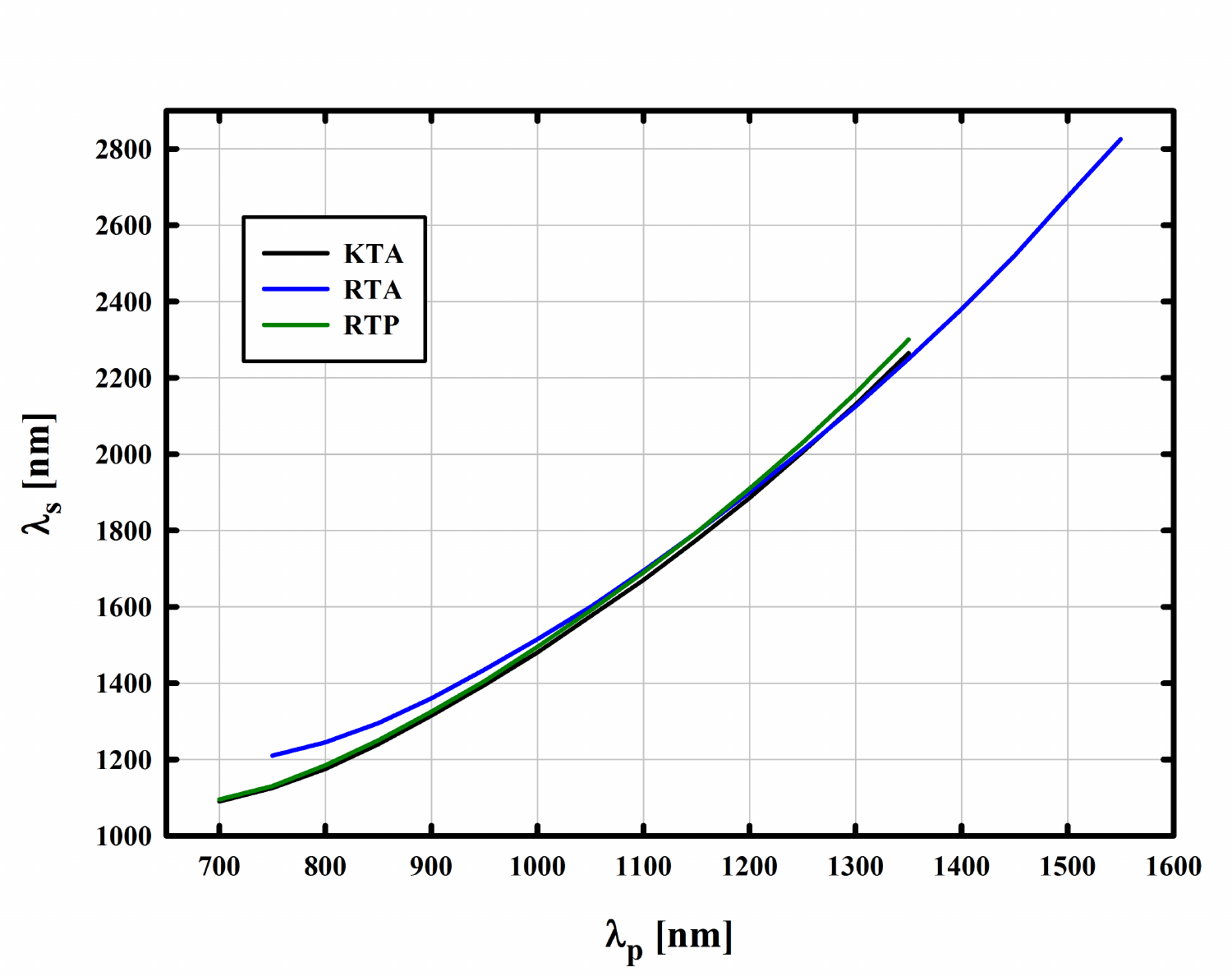}}
\caption{Pure type-II-SPDC states with bulk (a)~\text{KTP} and (b)~\text{KTA}, \text{RTA} and \text{RTP}. (Figure~(a) has been previously published in~\cite{laudenbach2016modelling}.)}
\label{KTPKTARTARTPnopoling}
\end{figure*}

As indicated by \Cref{KTP_ooe,CTA_ooe,KTA_ooe,RTA_ooe,RTP_ooe}, for each of the investigated materials there is a domain where intrinsically pure type-II SPDC can be achieved with crystals of infinite poling period $\Lambda$, i.e.\ for crystals which are not periodically poled at all. This brings of course the advantage of cost-efficient experimental solutions since the elaborate manufacturing of ferroelectric poling in micrometer domains becomes obsolete. The possible configurations allowing for this approach are depicted in \Cref{CTAnopoling,KTPKTARTARTPnopoling}.

For most setups the accessible signal and/or idler wavelengths are beyond $\SI{2}{\micro\metre}$, a regime which is\textemdash while invisible for up-to-date Si- or InGaAs photo diodes\textemdash currently approached by superconducting nanowire detectors~\cite{hadfield2016super, marsili2013mid}. However, as illustrated in \Cref{CTAnopoling}, CTA makes up a pleasant exception: According to our simulations, bulk (hence not periodically poled) CTA allows for the generation of intrinsically pure and frequency degenerate type-II SPDC in the popular telecom regime $\SI{775}{\nano\metre}~(o) \longrightarrow \SI{1550}{\nano\metre}~(o) +\SI{1550}{\nano\metre}~(e)$.

\section{Intrinsic Entanglement}

\begin{figure*}
	\centering
	\includegraphics[width=0.77\linewidth]{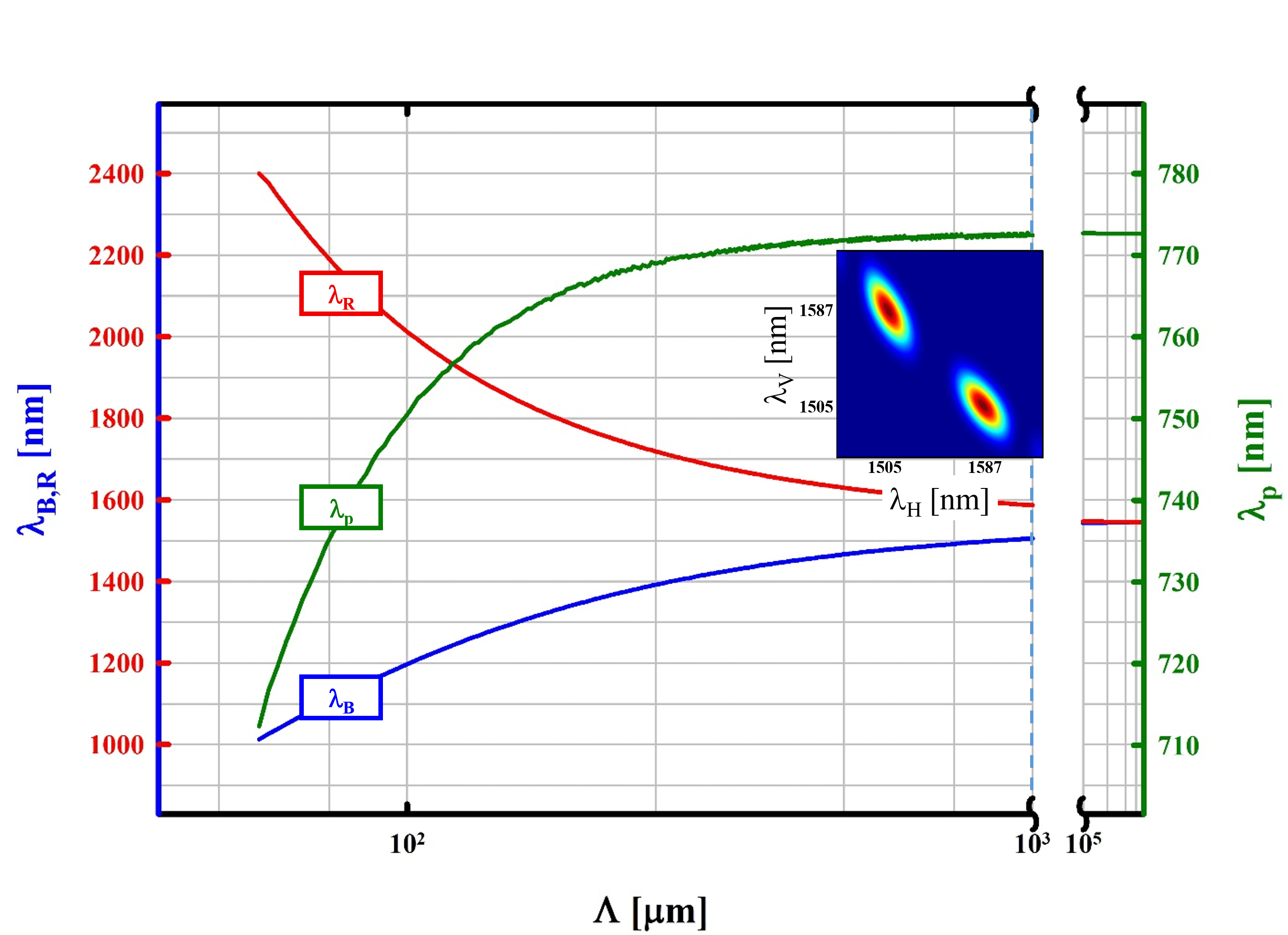}
\caption{ppCTA, $o \longrightarrow o + e$, intrinsic entanglement. Possible configurations allowing for intrinsic entanglement of photons generated by CDDC in \text{ppCTA}. The blue and red lines represent the short- and longwave photons respectively (each coming in both polarisations $H$ and $V$); the green line represents the wavelength of the pump laser. With increasing crystal periodicity $\Lambda$ the wavelengths $\lambda_{B}$ and $\lambda_{R}$ are approaching each other  which gives rise to the opportunity of creating photon pairs with similar bandwidths and group velocities. The vertical dashed light-blue line and the inset represent the downconversion $\SI{772.5}{\nano\metre}~(o) \longrightarrow \SI{1505}{\nano\metre}~(o,e) +\SI{1587}{\nano\metre}~(e,o)$, which is not only of interest due to the telecom wavelengths but also due to similar bandwidths and group delay in the crystal, allowing for high entanglement visibility and brightness (since no bandpass-filtering is required).}
\label{CTAentanglement}
\end{figure*}

\begin{figure*}
	\centering
\subcaptionbox{ppKTP}
	[0.49\linewidth]{\includegraphics[width=0.49\linewidth]{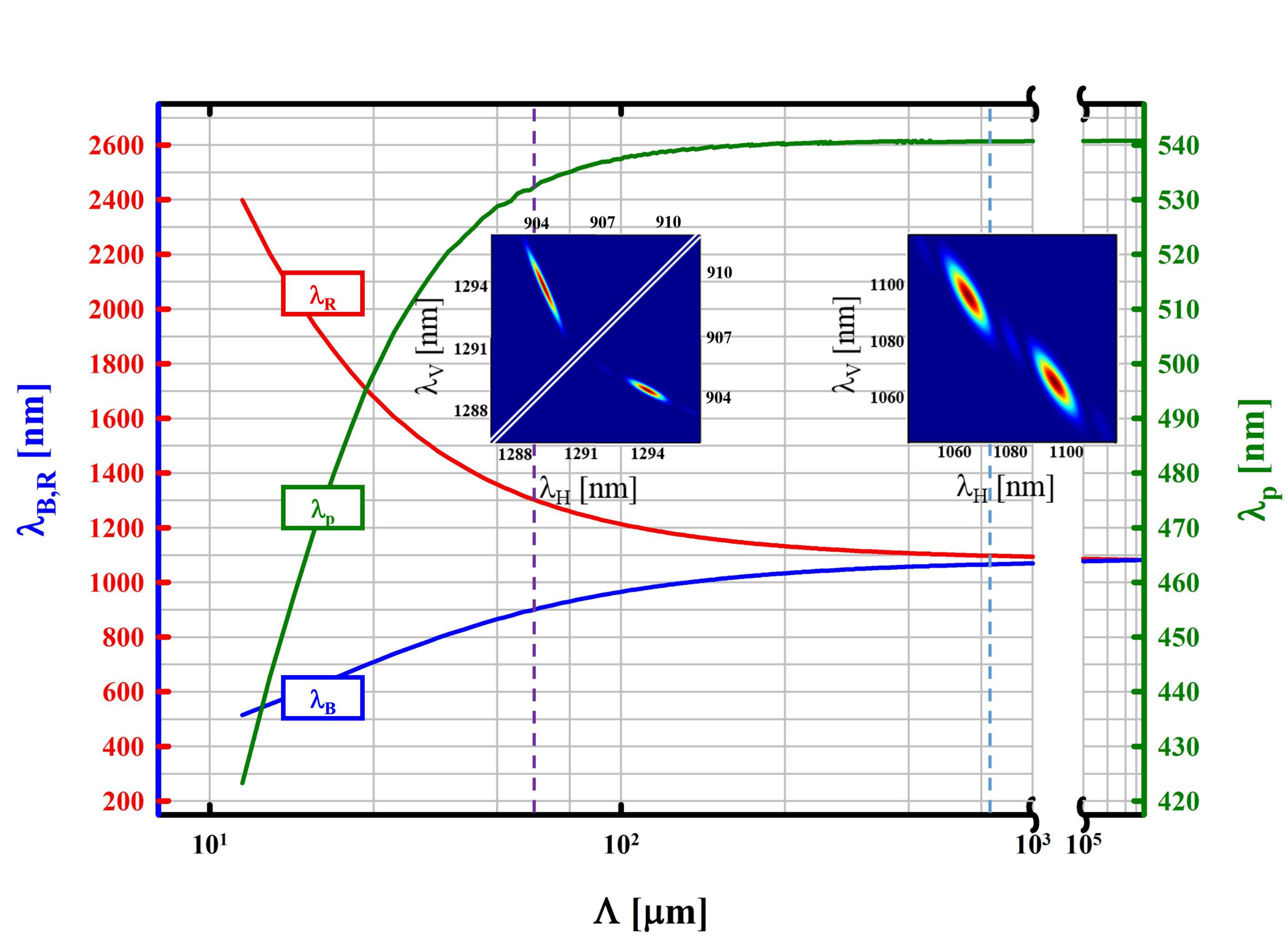}}
\subcaptionbox{ppKTA}
	[0.49\linewidth]{\includegraphics[width=0.49\linewidth]{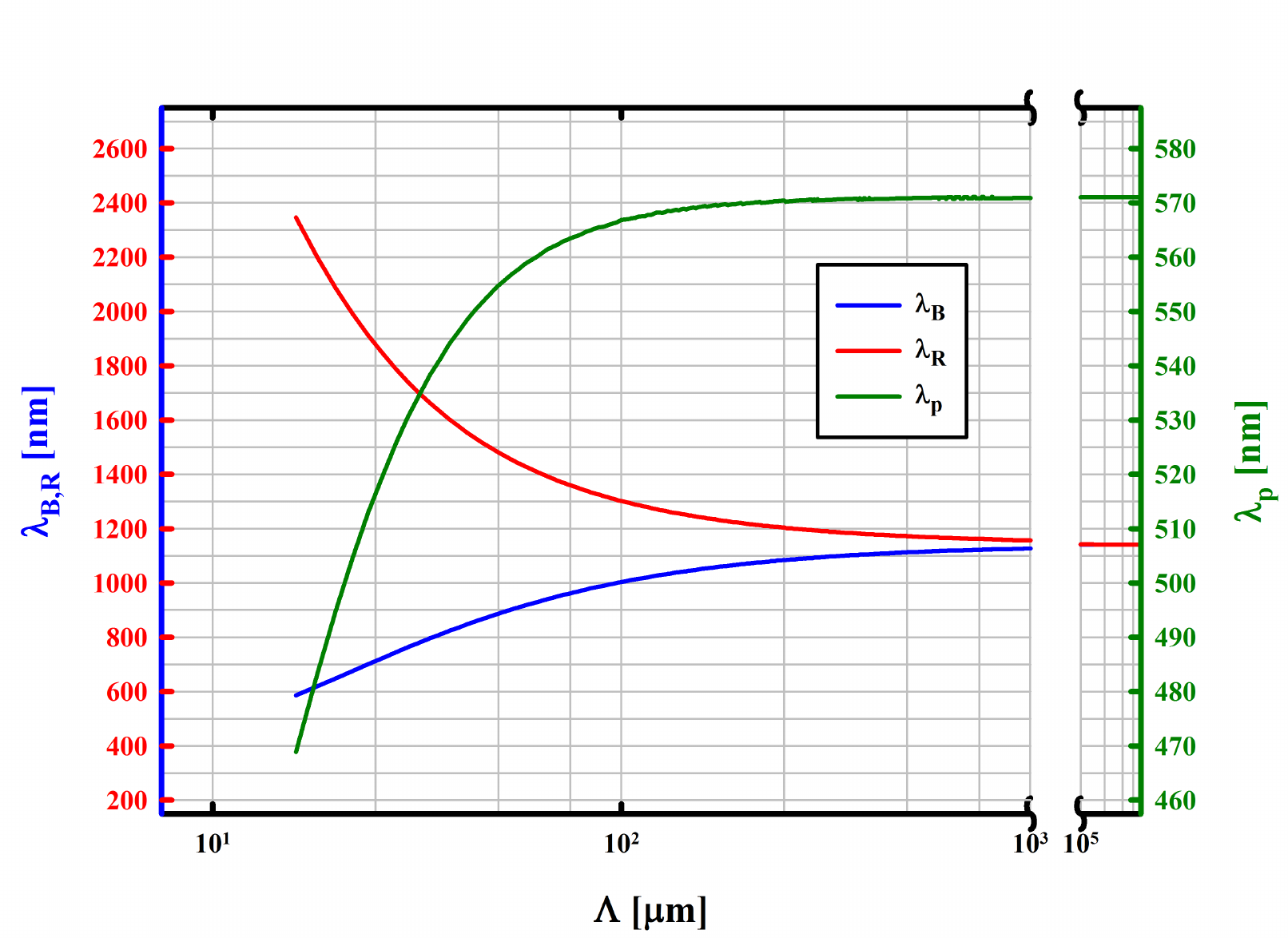}}
\subcaptionbox{ppRTA}
	[0.49\linewidth]{\includegraphics[width=0.49\linewidth]{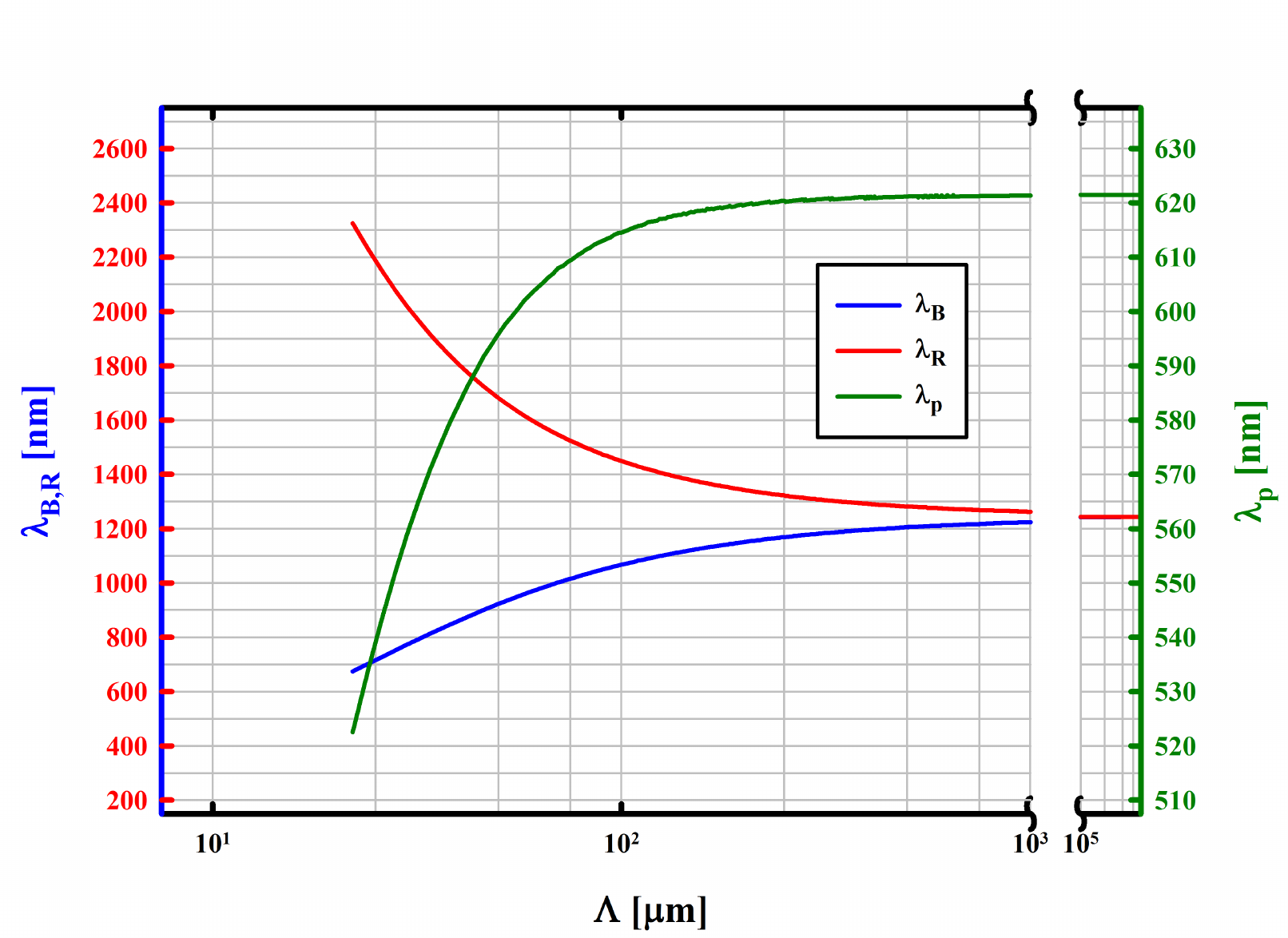}}
\subcaptionbox{ppRTP}
	[0.49\linewidth]{\includegraphics[width=0.49\linewidth]{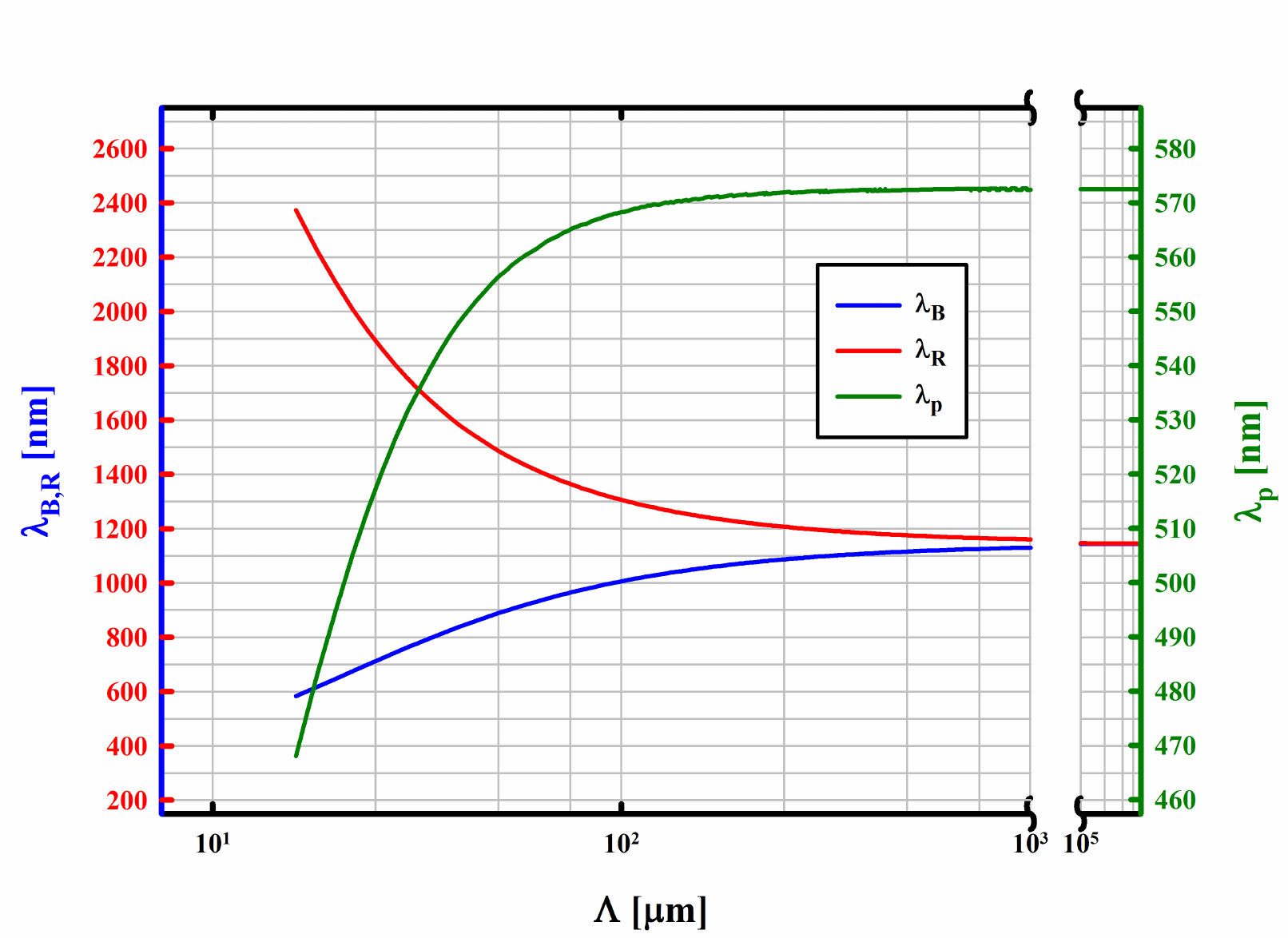}}
\caption{Possible wavelength configurations for intrinsic-entanglement generation in (a)~\text{ppKTP}, (b)~\text{ppKTA}, (c)~\text{ppRTA} and (d)~\text{ppRTP}. The vertical dashed purple line and corresponding inset graph in (a) highlight the first experimental demonstration $\lambda_{p0}=\SI{532.3}{\nano\metre}$, $\lambda_{B}=\SI{904.3}{\nano\metre}$ and $\lambda_{R}=\SI{1293.9}{\nano\metre}$ in \text{ppKTP}. The right inset graph over the vertical dashed light-blue line represents a configuration with similar bandwidth and group velocities. However, as these wavelengths lie around $\SI{1.1}{\micro \metre}$, they are difficult to detect with commonly used Si- or InGaAs detectors\textemdash a drawback which can be overcome by using \text{ppCTA}, as seen in \Cref{CTAentanglement}. (Figure (a) has been previously published in~\cite{laudenbach2016novel}.)}
\label{KTPKTARTARTPentanglement}
\end{figure*}

Until very recently the generation of entangled photon pairs using collinear SPDC required an elaborate setup, involving at least two superimposed beam paths~\cite{yoshizawa2003generation, clausen2014source, herbauts2013demonstration, pelton2004bright, huebel2007high, shi2004generation, koenig2005efficient, kim2006phase, fedrizzi2007wavelength, hentschel2009three} or a periodically poled crystal with alternating poling periodicities~\cite{suhara2009quasi, herrmann2013post}. In our recent paper~\cite{laudenbach2016novel} we demonstrated a novel method to generate photonic frequency- and polarisation entanglement, using only one unidirectionally pumped crystal with uniform poling period. Our method is based on the fact that a periodically poled nonlinear crystal with grating constant $\Lambda$, pumped by a laser with centre wavelength $\lambda_{p}$, can provide phase-matching for two different SPDC processes simultaneously, referred to as collinear double downconversion (CDDC). The poling period required to support a given downconversion configuration obeys

\begin{align}
\Lambda & = m \frac{2 \pi}{\Delta k} \notag \\
& = m \frac{2 \pi}{ k_{p} - k_{s} - k_{i}} .
\end{align}
In principle, two different pairs of signal and idler wavelength $\lambda_{s1}$, $\lambda_{i1}$ and $\lambda_{s2}$, $\lambda_{i2}$ can yield a momentum difference $\Delta k$ of equal magnitude but opposite sign. This will, according to the above equation, lead to the same periodicity $\Lambda$, however with one SPDC process exploiting the QPM order $m=1$ and for the other $m=-1$. Now, under careful choice of experimental parameters, two simultaneous downconversion processes can generate photon pairs with coinciding wavelengths but opposite polarisation:

\begin{subequations}
\begin{align}
\lambda_{H1}(m=1) & =\lambda_{V2}(m=-1), \\
\lambda_{V1}(m=1) & =\lambda_{H2}(m=-1),
\end{align}
\end{subequations}
where the corresponding wave numbers $k=2\pi n/\lambda$ fulfil the relation

\begin{align}
k_{H1}+k_{V1}=-(k_{H2}+k_{V2}) .
\end{align}
Considering that in each pair one photon is short-waved (blue, B) and the other one long-waved (red, R), we obtain the polarisation- and wavelength-entangled state

\begin{align}
\ket{\Psi} & = \alpha \ \hat{a}^{\dagger}_{HB} \hat{a}^{\dagger}_{VR} \ket{0} + \beta \ \hat{a}^{\dagger}_{HR} \hat{a}^{\dagger}_{VB} \ket{0} \notag \\
& \eqcolon \alpha \ket{H V} \otimes \ket{B R} + \beta \ket{H V}\otimes  \ket{R B} ,
\end{align}
which was successfully experimentally demonstrated in \cite{laudenbach2016novel}. This method is however only allowed for by particular nonlinear crystals, each supporting a finite set of possible wavelength configurations. \Cref{CTAentanglement,KTPKTARTARTPentanglement} illustrate possible experimental realisations of inherent-entanglement generation in the family of KTP-isomorphic crystals. For each poling period $\Lambda$ on the $x$-axis, there is one given pump wavelength $\lambda_{p}$ which generates daughter photons with wavelengths $\lambda_{B}$ and $\lambda_{R}$, each coming in horizontal and vertical polarisation. The graphs show that for long periods $\Lambda$ the signal and idler wavelength become more and more similar. This is beneficial for the entanglement visibility since relative differences of the group delays due to wavelength dispersion become neglectable. Furthermore, the left inset in \Cref{KTPKTARTARTPentanglement}~(a) illustrates that for smaller $\Lambda$ the two SPDC processes generate photons with interchangeable centre wavelengths but different spectral bandwidths. This introduces a certain degree of spectral distinguishability which undermines the entanglement visibility, an impairment which can be overcome by insertion of a bandpass filter in one channel. However, as suggested by the insets in \Cref{CTAentanglement} and \Cref{KTPKTARTARTPentanglement}~(a), for larger periods $\Lambda$ the bandwidths become very similar, therefore allowing for high indistinguishability even without bandpass filtering.

It turns out that, as illustrated in \Cref{CTAentanglement}, the material \text{ppCTA} offers the convenient opportunity to generate entangled photons with similar bandwidth (hence no filtering required) and low relative group delay ($( \Delta \text{GD})_{si} \approx \SI{0.2}{\pico\second / \milli\metre}$) with both photons in the telecom band.

\section{Conclusion}

In conclusion, we presented a profound numerical evaluation of five nonlinear materials, KTP (\ce{KTiOPO4}), CTA (\ce{CsTiOAsO4}), KTA (\ce{KTiOAsO4}), RTA (\ce{RbTiOAsO4}) and RTP (\ce{RbTiOPO4}), and the opportunities they offer in terms of spectrally pure downconversion states and entanglement generation. In particular, we found that CTA allows for the generation of collinear photon pairs with degenerate telecom wavelength $\SI{1550}{\nano\metre}$, orthogonal polarisation and high spectral purity at the same time\textemdash all this without the need for ferroelectric periodic poling. As a second remarkable feature of this material, we showed that periodically poled CTA can be used for the compact and efficient generation of polarisation- and frequency-entangled photon pairs with high visibility in the telecom regime. In addition, we numerically investigated the relation of purity and Hong-Ou-Mandel visibility and modelled the effects of bandpass filtering in terms of intensity loss and quantum performance.

We believe that this work can be a helpful resource to any experimentalist who uses periodically poled nonlinear crystals in order to generate highly-performing photon pairs. As we hope, this article will act as an incentive to further investigate and manufacture the more exotic nonlinear materials \text{CTA}, \text{KTA}, \text{RTA} and \text{RTP}.

\section{Acknowledgements}

This work was funded by the Austrian Research Promotion Agency (Österreichische Forschungsförderungsgesellschaft, FFG) through KVQ (No. 4642983). Furthermore, C.G. and Ph.W acknowledge support from the European Commission through EQUAM (No. 323714), PICQUE (No. 608062) and QUCHIP (No. 641039), as well as from the Austrian Science Fund (FWF) through START (Y585-N20), CoQuS (W1210-4) and NaMuG (P30067-N36).

\end{multicols}


\begin{thebibliography}{99}

\bibitem{meyer2017filtering} E. Meyer-Scott, N. Montaut, J. Tiedau, L. Sansoni, H. Herrmann, T. J. Bartley, and C. Silberhorn, Filtering is not enough for pure, efficient photon pairs, https://arxiv.org/abs/1702.05501 (2017).

\bibitem{Guerriero2013} T. Guerreiro, A. Martin, B. Sanguinetti, N. Bruno, H. Zbinden, and R. T. Thew, High efficiency coupling of photon pairs in practice, Opt. Express \textbf{21}, 27641--27651 (2013).

\bibitem{laudenbach2016modelling} F. Laudenbach, H. Hübel, M. Hentschel, P. Walther, and A. Poppe, Modelling parametric down-conversion yielding spectrally pure photon pairs, Opt. Express \textbf{24}, 2712--2727 (2016).

\bibitem{mosley2008recipe} P. J. Mosley, J. S. Lundeen, B. J. Smith, and I. A. Walmsley, Conditional preparation of single photons using parametric downconversion: a recipe for purity, New J. Phys. \textbf{10}, 093011 (2008).

\bibitem{uren2006generation} A. B. U'Ren, C. Silberhorn, R. Erdmann, K. Banaszek, W. P. Grice, I. A. Walmsley, and M. G. Raymer, Generation of pure-state single-photon wavepackets by conditional preparation based on spontaneous parametric downconversion, https://arxiv.org/abs/quant-ph/0611019 (2006).

\bibitem{jin2013widely} R.-B. Jin, R. Shimizu, K. Wakui, H. Benichi, and M Sasaki, Widely tunable single photon source with high purity at telecom wavelength, Opt. Express \textbf{21}, 10659--10666 (2013).

\bibitem{evans2010bright} P. G. Evans, R. S. Bennink, W. P. Grice, T. S. Humble, and J. Schaake, Bright source of spectrally uncorrelated polarization-entangled photons with nearly single-mode emission, Phys. Rev. Lett. \textbf{105}, 253601 (2010).

\bibitem{eckstein2011highly} A. Eckstein, A. Christ, P. J. Mosley, and C. Silberhorn, Highly efficient single-pass source of pulsed single-mode twin beams of light, Phys. Rev. Lett. \textbf{106}, 013603 (2011).

\bibitem{edamatsu2011photon} K. Edamatsu, R. Shimizu, W. Ueno, R.-B. Jin, F. Kaneda, M. Yabuno, H. Suzuki, S. Nagano, A. Syouji, and K. Suizu, Photon pair sources with controlled frequency correlation, Prog. Inform. \textbf{8}, 19--26 (2011).

\bibitem{yabuno2012fourphoton} M. Yabuno, R. Shimizu, Y. Mitsumori, H. Kosaka, and K. Edamatsu, Four-photon quantum interferometry at a telecom wavelength, Phys. Rev. A \textbf{86}, 010302 (2012).

\bibitem{jin2013nonclassical} R.-B. Jin, K. Wakui, R. Shimizu, H. Benichi, S. Miki, T. Yamashita, H. Terai, Z. Wang, M. Fujiwara, and M. Sasaki, Nonclassical interference between independent intrinsically pure single photons at telecommunication wavelength, Phys. Rev. A \textbf{87}, 063801 (2013).

\bibitem{jin2014efficient} R.-B. Jin, R. Shimizu, I. Morohashi, K. Wakui, M. Takeoka, S. Izumi, T. Sakamoto, M. Fujiwara, T. Yamashita, S. Miki, H. Terai, Z. Whang, and M. Sasaki, Efficient generation of twin photons at telecom wavelengths with 2.5 GHz repetition-rate-tunable comb laser, Sci. Rep. \textbf{4}, 7468 (2014).

\bibitem{weston2016efficient} M. M. Weston, H. M. Chrzanowski, S. Wollmann, A. Boston, J. Ho, L. K. Shalm, V. B. Verma, M. S. Allman, S. W. Nam, R. B. Patel, S. Slussarenko, and G. J. Pryde, Efficient and pure femtosecond-pulse-length source of polarization-entangled photons, Opt. Express \textbf{24}, 10869--10879 (2016).

\bibitem{grice2001eliminating} W. P. Grice, A. B. U’ren, and I. A. Walmsley, Eliminating frequency and space-time correlations in multiphoton states, Phys. Rev. A \textbf{64}, 063815 (2001).

\bibitem{mosley2008heralded} P. J. Mosley, J. S. Lundeen, B. J. Smith, P. Wasylczyk, A. B. U’Ren, C. Silberhorn, and I. A. Walmsley, Heralded generation of ultrafast single photons in pure quantum states, Phys. Rev. Lett. \textbf{100}, 133601 (2008).

\bibitem{Dixon2013} P. B. Dixon, J. H. Shapiro, and F. N. C. Wong, Spectral engineering by Gaussian phase-matching for quantum
photonics, Opt. Express \textbf{21}, 5879–5890 (2013).

\bibitem{Dosseva2016} A. Dosseva, \L. Cincio, and A. M. Bra\'{n}czyk, Shaping the joint spectrum of down-converted photons through
optimized custom poling, Phys. Rev. A \textbf{93}, 013801 (2016).

\bibitem{graffitti2017pure} F. Graffitti, D. Kundys, D. T. Reid, A. M. Bra\'{n}czyk, and A. Fedrizzi, Pure down-conversion photons through sub-coherence-length domain engineering, Quantum Sci. Technol. \textbf{2}, 035001 (2017).

\bibitem{Chen2017} C. Chen, C. Bo, M. Y. Niu, F. Xu, Z. Zhang, J. H. Shapiro, and F. N. C. Wong, Efficient generation and characterization of spectrally factorable biphotons, Opt. Express \textbf{25}, 7300--7312, (2017).

\bibitem{laudenbach2016novel} F. Laudenbach, S. Kalista, M. Hentschel, P. Walther, and H. Hübel, A novel single-crystal \& single-pass source for polarisation- and colour-entangled photon pairs, Sci. Rep. \textbf{7}, 7235 (2017).

\bibitem{yoshizawa2003generation} A. Yoshizawa, R. Kaji, and H. and Tsuchida, Generation of polarization-entangled photon pairs at 1550 nm using two PPLN waveguides, Electron. Lett. \textbf{39}, 621--622 (2003).

\bibitem{clausen2014source} C. Clausen, F. Bussières, A. Tiranov, H. Herrmann, C. Silberhorn, W. Sohler, M. Afzelius, and N. Gisin, A source of polarization-entangled photon pairs interfacing quantum memories with telecom photons, New J. Phys. \textbf{16}, 093058 (2014).

\bibitem{herbauts2013demonstration} I. Herbauts, B. Blauensteiner, A. Poppe, T. Jennewein, and H. Huebel, Demonstration of active routing of entanglement in a multi-user network, Opt. Express \textbf{21}, 29013--29024 (2013).

\bibitem{pelton2004bright} M. Pelton, P. Marsden, D. Ljunggren, M. Tengner, A. Karlsson, A. Fragemann, C. Canalias, and F. Laurell, Bright, single-spatial-mode source of frequency non-degenerate, polarization-entangled photon pairs using periodically poled KTP, Opt. Express \textbf{12}, 3573--3580 (2004).

\bibitem{huebel2007high} H. Hübel, M. R. Vanner, T. Lederer, B. Blauensteiner, T. Lorünser, A. Poppe, and A. Zeilinger, High-fidelity transmission of polarization encoded qubits from an entangled source over 100 km of fiber, Opt. Express \textbf{15}, 7853--7862 (2007).

\bibitem{shi2004generation} B. S. Shi and A. Tomita, Generation of a pulsed polarization entangled photon pair using a Sagnac interferometer, Phys. Rev. A \textbf{69}, 013803 (2004).

\bibitem{koenig2005efficient} F. König, E. J. Mason, F. N. C. Wong, and M. A. Albota, Efficient and spectrally bright source of polarization-entangled photons, Phys. Rev. A \textbf{71}, 033805 (2005).

\bibitem{kim2006phase} T. Kim, M. Fiorentino, and F. N. C. Wong, Phase-stable source of polarization-entangled photons using a polarization Sagnac interferometer, Phys. Rev. A \textbf{73}, 012316 (2006).

\bibitem{fedrizzi2007wavelength} A. Fedrizzi, T. Herbst, A. Poppe, T. Jennewein, and A. Zeilinger, A wavelength-tunable fiber-coupled source of narrowband entangled photons, Opt. Express \textbf{15}, 15377--15386 (2007).

\bibitem{hentschel2009three} M. Hentschel, H. Hübel, A. Poppe, and A. Zeilinger, Three-color Sagnac source of polarization-entangled photon pairs, Opt. Express \textbf{17}, 23153--23159 (2009).

\bibitem{horn2013inherent} R. T. Horn, P. Kolenderski, D. Kang, P. Abolghasem, C. Scarcella, A. Della Frera, A. Tosi, L. G. Helt, S. V. Zhukovsky, J. E. Sipe, G. Weihs, A. S. Helmy, and Thomas Jennewein, Inherent polarization entanglement generated from a monolithic semiconductor chip, Sci. Rep. \textbf{3}, 2314 (2013).

\bibitem{kang2016monolithic} D. Kang, A. Anirban, and A. S. Helmy, Monolithic semiconductor chips as a source for broadband wavelength-multiplexed polarization entangled photons, Opt. Express \textbf{24}, 15160--15170 (2016).

\bibitem{jin2016spectrally} R.-B. Jin, P. Zhao, P. Deng, and Q. L. Wu, Spectrally Pure States at Telecommunications Wavelengths from Periodically Poled M TiO X O 4 (M= K, Rb, Cs; X= P, As) Crystals, Phys. Rev. Applied \textbf{6}, 064017 (2016).

\bibitem{dmitriev1999handbook} V.G. Dmitriev, G. G. Gurzadyan, and D. N. Nikogosyan, \textit{Handbook of Nonlinear Optical Crystals}, 3rd ed. (Springer-Verlag, Berlin, 1999).

\bibitem{sands1993dover} D. E. Sands, \textit{Introduction to Crystallogrophy}, (Dover Publications, New York, 1993).

\bibitem{laudenbach2016qpmoptics} F. Laudenbach, H. Hübel, M. Hentschel, and A. Poppe, QPMoptics: a novel tool to simulate and optimise photon pair creation, in \textit{Proceedings of SPIE Photonics Europe 2016} (International Society for Optics and Photonics, Brussels, 2016), pp. 98940V.

\bibitem{asphotonics} A. Smith, SNLO, http://www.as-photonics.com/snlo.

\bibitem{powers2011fundamentals} P. E. Powers, \textit{Fundamentals of nonlinear optics}, 1st. ed. (CRC Press, Boca Raton, 2011).

\bibitem{hong1987measurement}  C. K. Hong, Z. Y. Ou, and L. Mandel, Measurement of subpicosecond time intervals between two photons by interference, Phys. Rev. Lett. \textbf{59}, 2044 (1987).

\bibitem{Kaneda2016} F. Kaneda, K. Garay-Palmett, A.B. U’Ren, and P. G. Kwiat, Heralded single-photon source utilizing highly nondegenerate, spectrally factorable spontaneous parametric downconversion, Opt. Express \textbf{24}, 10733--10747 (2016).

\bibitem{Gerrits2014} T. Gerrits, F. Marsili, V. Verma, L. Shalm, M. Shaw, R. Mirin and S. W.  Nam, Spectral Correlation Measurements at the Hong-Ou-Mandel Interference Dip, Phys. Rev. A \textbf{91}, 013830 (2014).

\bibitem{Jin2015swapping} R.-B. Jin, M. Takeoka, U. Takagi, R. Shimizu, and M. Sasaki, Highly efficient entanglement swapping and teleportation at telecom wavelength, Sci. Rep. \textbf{5}, 9333 (2015)

\bibitem{Grice2011} W. P. Grice, R. S. Bennink, D. S. Goodman, and A. T. Ryan, Spatial entanglement and optimal single-mode coupling, Phys. Rev. A, \textbf{83}, 023810 (2011).

\bibitem{Bruno2014} N. Bruno, A. Martin, T. Guerreiro, B. Sanguinetti, and R. T. Thew, Pulsed source of spectrally uncorrelated and indistinguishable photons at telecom wavelengths, Opt. Express \textbf{22}, 17246--17253 (2014).

\bibitem{Gajewski2016} A. Gajewski and P. Kolenderski, Spectral correlation control in down-converted photon pairs, Phys. Rev. A \textbf{94}, 013838 (2016).

\bibitem{Konig2004}
F.~K{\"o}nig and F.~N.~C. Wong, Extended phase matching of
  second-harmonic generation in periodically poled {KTiOPO}$_4$ with zero
  group-velocity mismatch, Appl. Phys. Lett. \textbf{84}, 1644 (2004).
  
\bibitem{Fradkin1999}
K.~Fradkin, A.~Arie, A.~Skliar, and G.~Rosenman, Tunable midinfrared
  source by difference frequency generation in bulk periodically poled
  {KTiOPO}$_4$, Appl. Phys. Lett. \textbf{74}, 914--916 (1999).

\bibitem{Kato2002}
K.~Kato and E.~Takaoka, Sellmeier and thermo-optic dispersion formulas
  for {KTP}, Appl. Opt. \textbf{41}, 5040--5044 (2002).

\bibitem{Mikami2009}
T.~Mikami, T.~Okamoto, and K.~Kato, Sellmeier and thermo-optic
  dispersion formulas for {RbTiOPO}$_4$, Opt. Mater. \textbf{31}, 1628--1630
  (2009).

\bibitem{Kato1994}
K.~Kato, Second-harmonic and sum-frequency generation in
  {KTiOAsO}$_4$, IEEE J. Quantum Electron. \textbf{30}, 881--883 (1994).

\bibitem{Cheng1993}
L.~T. Cheng, L.~K. Cheng, J.~D. Bierlein, and F.~C. Zumsteg, Nonlinear
  optical and electro-optical properties of single crystal {CsTiOAsO}$_4$,
  Appl. Phys. Lett. \textbf{63}, 2618--2620 (1993).  

\bibitem{Cheng1994}
L.~K. Cheng, L.~T. Cheng, J.~Galperin, P.~A.~M. Hotsenpiller, and J.~D.
  Bierlein, Crystal growth and characterization of {KTiOPO}$_4$
  isomorphs from the self-fluxes, J. Cryst. Growth \textbf{137}, 107--115
  (1994).

\bibitem{hadfield2016super} R. H. Hadfield and G. Johansson, \textit{Superconducting Devices in Quantum Optics} (Springer Verlag, Berlin, 2016).

\bibitem{marsili2013mid} F. Marsili, V. Verma, M. J. Stevens, J. A. Stern, M. D. Shaw, A. Miller, D. Schwarzer, A. Wodtke, R. P. Mirin, and S. W. Nam, Mid-infrared single-photon detection with tungsten silicide superconducting nanowires, in \textit{CLEO: Science and Innovations 2013} (Optical Society of America, San Jose, 2013), pp. CTu1H-1.

\bibitem{suhara2009quasi} T. Suhara, G. Nakaya, J. Kawashima, M. Fujimura, Quasi-phase-matched waveguide devices for generation of postselection-free polarization-entangled twin photons, IEEE Photonics Technol. Lett. \textbf{15}, 1096--1098 (2009).

\bibitem{herrmann2013post} H. Herrmann, X. Yang, A. Thomas, A. Poppe, W. Sohler, and C. Silberhorn, Post-selection free, integrated optical source of non-degenerate, polarization entangled photon pairs, Opt. Express \textbf{21}, 27981--27991 (2013).

\end{thebibliography}
\end{document}